\documentclass{aa}  

\usepackage{graphicx}
\usepackage{txfonts}

\usepackage{color, colortbl}
\usepackage[hidelinks]{hyperref}
\usepackage{xcolor}
\hypersetup{
    colorlinks = true, 
    linkcolor={red!50!black},
    citecolor={blue!50!black},
    urlcolor={blue!80!black}
}
\begin{document} 

\definecolor{bubbles}{rgb}{0.91, 1.0, 1.0}
\definecolor{columbiablue}{rgb}{0.61, 0.87, 1.0}
\definecolor{cream}{rgb}{1.0, 0.99, 0.82}
\definecolor{lightblue}{rgb}{0.68, 0.85, 0.9}
\definecolor{lightcyan}{rgb}{0.88, 1.0, 1.0}

   \title{More planetary candidates from K2 Campaign 5 by \textsc{\huge tran\_k\oldstylenums{2}}}

   \author{Geza Kovacs
          \inst{1}
          }

   \institute{Konkoly Observatory, Budapest, 1121 Konkoly Thege ut. 15-17, Hungary\\
              \email{kovacs@konkoly.hu}
             }

   \date{Received June 23, 2020; accepted ???? ??, 202?}


%
%
  \abstract
{The exquisite precision of the space-based photometric surveys and the unavoidable 
presence of instrumental systematics and intrinsic stellar variability call for the 
development of sophisticated methods that separate these signal components from those 
caused by planetary transits.}
{Here we introduce \textsc{tran\_k\oldstylenums{2}} a stand-alone Fortran code to 
search for planetary transits under the colored noise of stellar variability and 
instrumental effects. With this code we perform a survey for new candidates.}
{Stellar variability is represented by a Fourier series, and, if needed, by an 
autoregressive model to avoid excessive Gibbs overshoots at the edges. For the 
treatment of systematics, cotrending and external parameter decorrelation are 
employed by using cotrending stars with low stellar variability, the chip position 
and the background flux level at the target. The filtering is made within the 
framework of the standard weighted least squares, where the weights are determined 
iteratively, to allow robust fit and separate the transit signal from stellar 
variability and systematics. Once the periods of the transit components are 
determined from the filtered data by the box-fitting least squares method, we 
reconstruct the full signal and determine the transit parameters with a higher 
accuracy. This step greatly reduces the excessive attenuation of the transit 
depths and minimizes shape deformation.} 
{The code was tested on the field of Campaign 5 of the K2 mission. We detected 
$98$\% of the systems with all their candidate planets reported earlier by other 
authors, surveyed the whole field and discovered $15$ new systems. Additional 
$3$ planets were found in $3$ multiplanetary systems and $2$ more planets were 
found in a previously known single planet system.}
   {}


\titlerunning{More planetary candidates from K2 Campaign 5}
\authorrunning{Kovacs, G.}

   \maketitle
%
%
%
\section{Introduction}
\label{sect:intro}
Tackling instrumental systematics (colored noise) has been a major issue 
from the very early years of the wide-field ground-based surveys (e.g., 
TrES, SuperWASP, HATNet, XO). These systematics were very severe due to 
the large field of views, the ground-based nature of the observations 
and other imperfections attributed to the optics or CCD cameras used 
\citep{alonso2004,bakos2004,mccullough2005,pollacco2006}. Novel methods 
that were able to cure the crippling effect of systematics started to gain 
ground only a few years after these major projects started. These methods 
are based on the simple observation that systematics (by the very meaning 
of this word) should be common among many stars in the field, and therefore, 
can be used for correcting the target star of our interest.\footnote{By 
following the jargon of the developers of the Kepler post-processing 
pipeline \citep[see][]{smith2012,stumpe2012}, we often use the word 
``cotrending'' for this filtering process.} The Trend Filtering Algorithm 
\citep[TFA,][]{kovacs2005}, and Systematics Removal \citep[SysRem,][]{tamuz2005} 
use this idea. TFA is using it in a ``brute force'' way (with large 
number of correcting stars), whereas SysRem employs a more sophisticated 
approach (akin to Principal Component Analysis - PCA). With the aid of PCA,  
only the dominating systematics are included in the correction process 
for a given target. 

Additional effects hindering shallow signal discoveries have also been 
considered. Unlike commonalities in the flux changes of the objects in 
the field, differences in the pixel properties are uncommon, and therefore, 
curable only on a target-by-target basis. External Parameter Decorrelation 
(EPD) can mitigate the dependence on pixel sensitivity by including 
polynomials of the stellar image parameters (e.g., centroid position, 
size of the Point Spread Function - PSF, etc.) in the time series modeling \citep{knutson2008, bakos2010}. 
Intrinsic/physical variation of the stellar flux has become a more 
fundamental question with the advance of space missions. The first 
attempt to deal with this issue within the environment of systematics 
was presented by \cite{alapini2009}, within the context of the CoRoT 
mission. In their method, the stellar variation acts as a multiplicative 
noise source and searched for by an iterative method while fitting the 
raw stellar flux.    

The start of the Kepler mission in 2009 and later its very successful 
conversion to the two-wheel (K2) program (forced by the failure of the 
reaction wheels), together with the followup programs on additional space 
facilities, boosted further efforts in making transit search more efficient. 
The official post-processing pipeline (Presearch Data Conditioning -- PDC) 
uses PCA-selected basis vectors in a Bayesian framework to avoid overcorrection 
\citep{smith2012,stumpe2012}. In the early phase of the K2 mission, 
\cite{vanderburg2014} introduced the idea of Self Flat Filtering, SFF, 
based on the recognition of the tight correlation between a nonlinear 
combination of the image position and the observed stellar flux (akin 
to EPD). Although the method proved to be very successful, because of 
the absence of stellar variability in the model, other methods have 
also been developed to include both satellite roll correction and stellar 
variability. The method developed by \cite{aigrain2016} successfully 
tackle the issue by using a Gaussian Process (GP) model for the stellar 
variation. Yet another method, based on the idea of Pixel-level 
Decorrelation of \cite{deming2015}, was developed by \cite{luger2016}. 
It stands out from the other, `more traditional' models. Called as 
EVEREST, it utilizes only the pixel fluxes belonging to the target 
(no cotrending, no EPD). The assumption is that the incoming target 
flux is the same, whereas the pixel sensitivities are different and 
therefore, they can be transformed out from the total stellar flux. 
With the combination of GP modeling for stellar variability, it seems 
that this approach is very efficient in searching for transits. 
By applying EVEREST on the K2 Campaigns $0-8$, \cite{kruse2019} 
discovered $374$ new candidates, thereby nearly doubling the number 
of potential planets in these fields. Therefore, the method presented 
in this paper heavily relies on the sample on Campaign 5 (C05) 
of \cite{kruse2019} to perform a sanity check.        

Several other methods have been developed with similarities to the 
ones briefly described above. We refer the interested reader to a 
more complete list of methods in \cite{kovacs2017}. 

The purpose of this paper is to examine the possibility of further  
improvements in the K2 transit search methodologies.\footnote{The  
source code \textsc{tran\_k\oldstylenums{2}.f} developed in this 
work is available at \url{http://www.konkoly.hu/staff/kovacs/tran\_k2.html}, 
or, can be requested from the author.} Our approach is based on 
allowing large TFA template size, using Fourier series to represent 
stellar variability and protecting the transit signal by using a 
robust least squares method to perform the simultaneous TFA$+$Fourier 
filtering before the transit period search \citep[Box-fitting Least Squares -- BLS,][]{kovacs2002}.

%
%
\section{Datasets}
\label{sect:data}
The data analysis method presented in this paper is a post-processing 
step after the derivation of the time-dependent net fluxes from the 
images taken by the telescope. We do not deal with the various 
possibilities in getting these raw (Simple Aperture Photometry, SAP) 
fluxes. As a result, we rely on those datasets that have already been 
created by teams dealing with this demanding task. 

The raw fluxes used in this paper come from two sources: 
a) the official image reduction and post processing pipeline of the 
Kepler mission \citep{smith2012}, and 
b) the {\sc k\oldstylenums{2}phot}\footnote{\url{https://github.com/petigura/k2phot}} 
code, performing the aperture photometry 
in supplying the input time series for the TERRA pipeline \citep{petigura2012,petigura2015, aigrain2016}. 
These data are accessible via the NASA Exoplanet Archive\footnote{ 
\url{https://exoplanetarchive.ipac.caltech.edu/}} 
and the affiliated Exoplanet Follow-up Observing Program (ExoFOP)\footnote{
\url{https://exofop.ipac.caltech.edu/}}, and referred to in the following, respectively, 
as KEP and PET. We note that the latter set includes all available epochs 
in the campaigns (yielding an overall data point number of $\sim 3620$ 
per target), whereas the KEP set, to avoid instrumental transients, 
discards certain data items. This leaves $\sim 3430$ data points for 
the analysis.

%
%
\section{\textsc{tran\_k\oldstylenums{2}}: Overall description}
\label{sect:tran_k2}
In constructing a code with the ability of treating systematics 
and stellar variability without destroying the rare and shallow 
transit events, we considered four vital goals to follow: 
\begin{itemize}
\item
Use a nearly complete time series model by including internal 
(i.e., non-instrumental) variability and systematics in the filter 
acting on the input time series at the data preparation phase 
(before signal search).  
\item
Employ a wealth of cotrending field stars and image parameters 
following essentially the original ideas of TFA and EPD. 
\item
For the protection of the transit events, no data clipping 
should be used. Instead, employ robust fits on the original 
(raw) data with iterative weight adjustment.
\item
Once the transit periods are found, use a full model for 
signal reconstruction to compensate for the transit depression 
because of the use of an incomplete model at the signal search 
phase.      
\end{itemize}

According to these guidelines, by using broadly the notation of 
\cite{kovacs2018}, the main steps of data processing are as follows.    

\begin{itemize}
\item[1.]
We select $N_{\rm TFA}$ time series  
\begin{eqnarray} 
\{U_{\rm j}(i);\ j=1,2, ..., N_{\rm TFA};\ i=1,2, ..., N_{\rm j}\}\ , 
\end{eqnarray} 
from the field by employing the criteria to be discussed in 
Sect.~\ref{sect:tfa}. These time series are assumed to represent 
the commonality shared by most of the stars in the field. In the 
particular case of the Kepler mission, the number of data points 
per star, $N_{\rm j}$ is almost the same, likewise their time 
distributions. Therefore, the requested interpolation to the same 
timebase as that of the target star (as needed for the TFA filtering) 
is very safe.    
\item[2.]
Generate self-correcting time series from the background flux \{$B(i)$\} 
and the centroid pixel position \{$(X(i),Y(i))$\} of the target 
with $N$ datapoints   
\begin{eqnarray} 
\{B_{\rm j}(i);\ j=1,2,3;\ i=1,2, ..., N\}\ ,\\
\{Z_{\rm j}(i);\ j=1,2, ..., 9;\ i=1,2, ..., N\}\ ,
\end{eqnarray} 
where $B_1, B_2, B_3=B, B^2, B^3$, and 
$Z_1, Z_2, ...=X, Y, X^2, XY, Y^2, X^3, X^2Y, XY^2, Y^3$.  
These time series are scaled to unity and then zero-averaged.  
For \{$B(i)$\} a robust outlier-correction\footnote{Implying 
robust polynomial fit with Cauchy weights and iterative $3\sigma$ 
clipping.} is employed to avoid unwanted fluxes from neighboring 
stars.   
\item[3.]
Stellar variability in the target time series is represented by 
a set of sine and cosine functions   
\begin{eqnarray} 
\{S_{\rm k}(i), C_{\rm k}(i);\ k=1,2, ..., N_{\rm FOUR};\ i=1,2, ..., N\}\ , 
\end{eqnarray} 
where $N_{\rm FOUR}$ is the number of Fourier components.  
The frequencies are given by $kf_0$, where 
$f_0$ is close to the reciprocal of the total timebase. See 
Sects.~\ref{sect:detune} and \ref{sect:four_opt} for the choice 
of these parameters.  
\item[4.] 
Employ robust least squares minimization to determine the best 
fitting linear combination of the three signal constituents above 
to the target time series \{$T(i)$\} 
\begin{eqnarray}
\label{eq:D}
\mathcal D = \sum_{i=1}^N w(i)[T(i) - F(i)]^2 \ ,
\end{eqnarray}
with: 
\begin{eqnarray}
\label{eq:F}
F(i) = a_0 + \sum_{j=1}^{N_{\rm TFA}}a_{\rm j} U_{\rm j}(i) + 
\sum_{j=1}^{3}b_{\rm j} B_{\rm j}(i) + \sum_{j=1}^{9}c_{\rm j} Z_{\rm j}(i) + \nonumber \\
           \sum_{k=1}^{N_{\rm FOUR}}d_{\rm k} S_{\rm k}(i) + e_{\rm k} C_{\rm k}(i) \ .
\end{eqnarray}  
The weights \{$w(i)$\} are determined iteratively, starting with 
uniform weighting. We choose the Cauchy weight function   
\begin{eqnarray}
\label{eq:w}
w(i)={\sigma^2 \over \sigma^2+\Delta^2(i)} \ ,
\end{eqnarray}  
where $\sigma$ is the standard deviation of the fit and $\Delta(i)$ 
is the difference between the observed and the predicted values, i.e., 
$\Delta(i)=T(i) - F(i)$. At each step of the iteration we can solve 
the linear problem within the framework of standard weighted least 
squares, but then a correction is needed to the weights according to 
Eq.~\ref{eq:w}. Iteration is stopped when the relative change in 
$\sigma$ becomes less than $0.1$\%.  
\item[5.]
Minimize the effect of Gibbs overshooting\footnote{The Gibbs 
phenomenon is an asymptotic property of the Fourier decomposition, and 
follows from the `almost everywhere' type of convergence for square-integrable 
functions -- e.g., for continuous functions defined on finite intervals. 
It is often exhibited as high-frequency oscillations or overshootings if 
the order of the Fourier decomposition is high enough.} at the edges. This is done 
by fitting an autoregressive (AR) model\footnote{AR modeling 
performs backward/forward prediction from the linear combination of certain 
number of future/past values of a time series 
\citep[see Sect.~\ref{sect:ar} and][for early, and recent applications.]{fahlman1982,caceres2019}} 
to the inner part of the 
noise-mixed Fourier signal and predicting the values at the edges. 
The AR model is not incorporated in the more extended and full model 
fits described in step 7, but only used to prepare the data for the 
BLS analysis. We refer to Sect.~\ref{sect:ar} for further details on 
the AR edge correction.   
\item[6.]
With the converged regression coefficients, and considering possible 
change at the edges found at step 5, compute the residuals 
$\{R(i)=T(i)-F(i); i=1,2, ..., N\}$ and perform standard BLS transit 
search on $\{R(i)\}$ with successive prewhitening by the dominant 
BLS component found at each step of the prewhitening process. 
\item[7.]
After finding all significant transit components, supply the model 
given by Eq.~\ref{eq:F} with the transit model and compute the 
best transit depths and all other parameters entering in Eq.~\ref{eq:F} 
with this full model. Subtract the non-transiting part of the fit 
from the input signal $\{T(i)\}$ and derive new transit parameters 
from the residuals, allowing to fit all transit parameters (not only 
the depth, as in the previous step). With the new sets of transit 
parameters loop back to the full model and compute the next approximation 
for the transit depth (and for all other parameters, including the 
weights \{$w(i)$\} -- with the concomitant sub-iterations). The process 
is repeated until the same $\sigma$ condition is satisfied as mentioned 
at step 4. For the obvious time constraining nature, in the case of 
multiplanetary signals all components are treated separately when 
estimating the transit parameters of the individual components. However, 
the transit depths are different from this respect, since they can be 
estimated in a single grand linear fit as described above. We refer to step 7 as {\em signal reconstruction}, since during 
this step all constituents of the signal are considered and this 
leads to a (usually much) better approximation of the transit signal. 
\end{itemize}

In addition to the most crucial ingredients of the data analysis 
detailed above, there are many other, perhaps less crucial, but 
still important particularities worth mentioning. Some of them (e.g., 
matrix operation, iteration initialization) are related to the 
speed, others to the quality of the performance of the code. 
Because these latter ones have an effect on the detection efficacy, 
we briefly describe them below.  

\begin{figure}[h]
\centering
\includegraphics[width=0.40\textwidth]{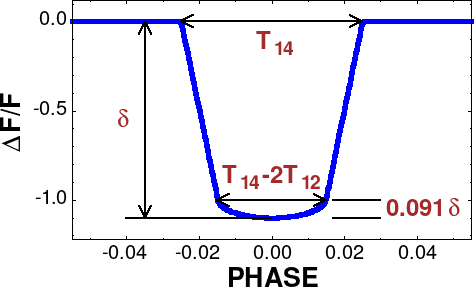}
\caption{Simplified transit shape used in this paper. The flux 
drops linearly in the ingress and egress parts at a rate given 
by the $T_{14}/T_{12}$ ratio and $\delta$. The U-shaped bottom 
is supposed to model the overall limb darkening. It has an 
analytical form of $u(x)=\delta-0.1(1-\sqrt{1-x^2})$, where 
$x=2\phi/(\phi_3-\phi_2)$, with totality start/end phases 
$\phi_2$, $\phi_3$ and phase $\phi$, where the function is to 
be evaluated.}
\label{utran}
\end{figure}

{\em Transit shape -} The assumption of box-like transit shape is used 
only in the period search. Because of the high precision of the Kepler 
data, we found it obligatory to use a better approximation of the 
transit (otherwise, when searching for additional transits, we may 
not be able to remove the already identified component in its full 
extent). Therefore, the transit shapes are assumed to be of trapezoidal, 
with a rounded bottom parts. To be more specific, we show the shape 
of this simplified transit model for a specific set of parameters in 
Fig.~\ref{utran}. In modeling a given transit, we adjust four parameters: 
$T_{\rm c}$, the center of the transit, $T_{14}$, the full transit 
duration, $T_{12}$, the ingress duration and $\delta$ the total transit 
depth. We do not adjust the relative depth of the rounded bottom (this 
is left always at $10$\% of the depth where the ingress ends).     
%

%
%

{\em Single outliers -} We recall that no outlier selection is made 
in the data processing up to the pre-BLS phase. Since the BLS search 
is sensitive to outliers, we focus on these, but carefully avoid 
cases when transit-like features are suspected. As a result, we 
consider only those instances when the flux decrease is limited 
to the middle point among three successive points. The outlier 
status is determined on the basis of the deviation from the average 
of the two fluxes left and right of the point of interest. If this 
deviation exceeds $3\sigma_0$ (with $\sigma_0$ being the sigma-clipped 
standard deviation of the pre-BLS phase time series), then the flux 
of this point is replaced by the flux average of the neighboring 
points. A similar procedure can be employed in the reconstruction 
phase, when the transit signal helps further identifying the single 
outliers. 

{\em Flare correction -}
In a non-negligible number of stars, flares may also jeopardize 
the success of the search for periodic transits, since these events 
increase the colored-noise component of the BLS spectrum. Here 
we resorted the simplest and far from optimum way to remedy the 
effect of flares. By considering only events of flux increase 
relative to the average or to a given transit signal -- depending 
on the stage of the analysis -- we replace the ``positive'' outliers 
by the average or the corresponding values of the transit signal.  
We do not correct any ``flare'' that does not exceed the $3\sigma_0$ 
limit as given above. 

%
%
\subsection{Detuned Fourier fit}
\label{sect:detune}
To handle stellar variability, we opted for traditional Fourier 
decomposition, whereby -- based on very fundamental and well-known 
properties of Fourier series defined on a finite interval -- we can 
fit the smooth part of the observed flux variation. Even though any 
stellar variability has an inherently stochastic component (primarily 
due to surface convection) the adjacent fluxes are correlated at 
some degree, and the variation appears to be continuous (albeit 
non-periodic). Consequently, we can apply Fourier decomposition 
for a very large class of variations. Even though this is true, 
we have to be aware of the fact that the finite timebase inevitably 
introduces discontinuities, and as such, induce what is commonly 
known Gibbs oscillations close to the edges of the 
dataset. It is 
obvious that one needs to remedy this problem, since these 
high-frequency oscillations can easily dominate the time series 
after corrections for systematics and stellar variability. 

To avoid any data loss by simply chopping off some fraction of both 
ends of the dataset, we searched for various methods in the literature 
devoted to Gibbs oscillation minimization (e.g., by using the Lanczos 
sigma factor\footnote{See Weisstein, Eric W. "Lanczos sigma Factor." 
From MathWorld--A Wolfram Web Resource.\\  
\url{https://mathworld.wolfram.com/LanczosSigmaFactor.html}}). 
Because we did not find a suitable one, we tested the idea of 
period detuning (a method we did not find in the literature scanned). 
Although the precise (or at least better than presented) discussion 
of the method is out of the scope of this paper, we illustrate 
the method at work on an example. The idea is simple. Increase 
the fundamental period (and its harmonics) by some amount and 
thereby push the oscillations outside the timebase of interest. 

\begin{figure}[h]
\centering
\includegraphics[width=0.40\textwidth]{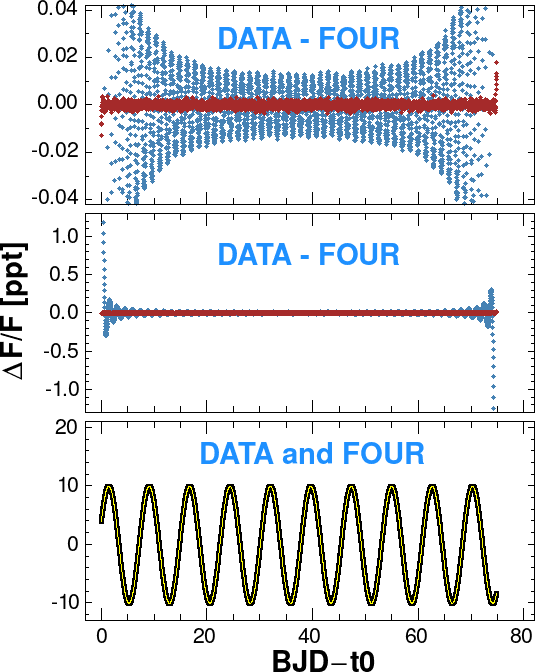}
\caption{Illustration of the effect of frequency detuning with 
fixed frequency components \{$f_i=(i+1)/(\beta\times T)$\}, where 
$T$ is the total time span and $\beta=1.0$ for the classical 
Fourier decomposition (steelblue dots) and $\beta=1.05$ for the 
detuned case (brown dots). The lower panel shows the input 
sinusoidal signal (black line) and the Fourier fit 
(yellow line -- there is no $\beta$ dependence on this scale)}. 
\label{detune}
\end{figure}

We show the effect of detuning in Fig.~\ref{detune}. We generated 
a sinusoidal signal on the observed timebase of one of the members 
of Campaign 5. We chose the following parameters: $P=7.6508$~d, 
$A=10$~ppt (parts per thousand, i.e., $A=0.010$), with an additive Gaussian (white) noise of 
$\sigma=0.001$~ppt. We set the order of the Fourier fit equal to 
$50$. We see that detuning works very efficiently, with only a 
negligible overshooting close to the noise level. The degree 
by which detuning may eliminate Gibbs phenomenon depends on the 
type of signal and the order of the Fourier series. Nevertheless, 
the method worked neatly in nearly all cases encountered so far 
during the analysis of the $20000$ stars of Campaign 5.    

%
%
%

%
%
\subsection{Autoregressive edge correction}
\label{sect:ar}
Here we focus on those (not very numerous) cases when detuning 
does not work. Instead of trying to adjust the Fourier order, we 
opted to employ an AR model on the noise-perturbed Fourier fit 
and predict the Gibbs-free signal at the edges. The AR modeling 
is always performed and compared with the Fourier fit and the one 
is chosen that yields a better fit to the data. The main 
steps of the AR modeling are as follows.  

\begin{itemize}  
\item[a.]
Cut some fraction ($L$ data points, some $5$\% the full set of 
$N$ points) of the Fourier signal at the edges of the dataset. 
Add Gaussian noise to the signal to avoid complete fit to 
the Fourier solution and thereby reproducing the Gibbs oscillations 
at the edges, where the AR model is used to predict the oscillation-free 
continuation of the Fourier solution. The size of the noise can 
be estimated from the fit based on the Fourier model. 
\item[b.]
Interpolate the above noisy Fourier signal \{$x(i); i=1,2, ..., n$\} 
to an equidistant timebase and fit it by a high-order AR model. 
To maintain stability we use an AR order of $m=900$. To determine 
the autoregressive coefficients \{$a_{j}$\}, we use one-side  
predictions for the outermost $m$ data points, i.e.,

\begin{eqnarray}
\label{ar_1}
x(k) = \sum_{i=1}^m a_{i}x(k+i) 
\ \ \ {\rm and} \ \ \   
x(k) = \sum_{i=1}^m a_{i}x(k-i)\ ,
\end{eqnarray}
for the left and the right ends, respectively. For the middle of the 
dataset (i.e., from $k=m+1$ to $k=n-m$ we use a two-side model with 
simple arithmetic mean 
\begin{eqnarray}
\label{ar_2}
x(k) = {1 \over 2} \sum_{i=1}^m a_{i}[x(k+i) + x(k-i)]\ .
\end{eqnarray}
Least squares with equal weights are used to determine the AR 
coefficients \{$a_{i}$\}. 
\item[c.]
Predict the outermost $L$ values of the Fourier points from those 
interior to the edges. We found that these predictions may still 
have some curvature or mild wavy behavior. To eliminate this feature, 
and protect any possible transit feature close to the end points, 
we robustly fit $5^{\rm th}$-order polynomials to the AR predictions.    
\item[d.]
Compare the residuals between the systematics-free data and 
approximations obtained by the AR and Fourier modeling. Select 
the one that yields smaller Root Mean Square (RMS) and pass the so-obtained 
systematics- and variability-free time series to the BLS routine.   
\end{itemize}  

\begin{figure}[h]
\centering
\includegraphics[width=0.40\textwidth]{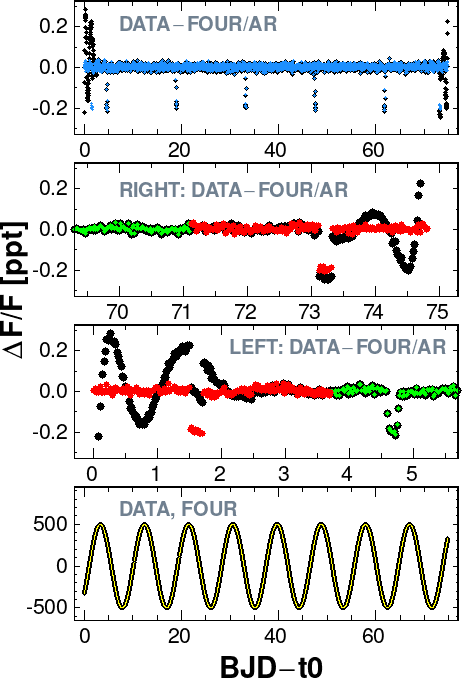}
\caption{Illustration of the elimination of the Gibbs oscillations 
at the edges of the dataset by using AR modeling. 
{\em Bottom:} input dataset (black line: noisy sinusoidal with 
transits), Fourier fit (yellow). The transits are not visible at 
this scale. 
{\em Middle panels:} residual signal after subtracting the best 
Fourier fit and the best AR model (black and red dots, respectively). 
The green dots indicate where the Fourier fit was accepted. 
{\em Uppermost panel:} residual time series obtained by Fourier 
(black) and AR modeling (light blue). See text for the main signal parameters.} 
\label{ar_model}
\end{figure}

To illustrate the method at work, we generated a synthetic sinusoidal 
signal with $P=9.091$~d, semi-amplitude $A=500$~ppt and superposed a 
transit signal with $P_{\rm tr}=14.286$~d, $\delta=0.2$~ppt. We also 
added two individual boxy transits near the end points to test if the 
method will preserve these events. The order of the Fourier series was 
$50$. The result is shown in Fig.~\ref{ar_model}. It is important to 
note that the Fourier fit was made with detuned fundamental period 
(see Sect.~\ref{sect:detune}). As it was emphasized earlier, detuning 
is very efficient in the elimination of the Gibbs oscillations. If 
detuning is not efficient enough, AR modeling may go to the rescue 
and allow the detection of faint signals.

%
%
%

%
%
\subsection{TFA template selection}
\label{sect:tfa}
The selection of the proper set of cotrending time series is one 
of the fundamental questions in filtering out systematics from the 
observed signal. The most liberal choice one can try is to avoid clear variables 
showing excessive scatter, but otherwise paying not too much attention 
to stellar variability in the cotrending set \citep[i.e.,][]{kovacs2005}. 
The somewhat vague (but -- under certain circumstance -- working) 
argument here is that in the presence of noise and targeting transits, 
the variability of the cotrending time series will be scaled down 
due to the nearly constant nature of the target time series. The 
transits, due to their short time scales will not be much effected, 
and the final result will be positive, since the cotrending time 
series will fit primarily the systematics, due to their coherence 
with these components in the flux variation of the target. Interestingly, 
this approach has been fairly successful over the numerous applications 
in searching for periodic transits. 

In spite of the success of the above `brute force' method, efforts 
have been made to decrease the number of cotrending time series 
as much as possible, to: 
a) avoid signal degradation due to overfitting, 
b) using only the essential (i.e., `most common') components and 
aid a better separation of the transit signal and systematics. 
The most obvious approach in reaching these goals is to employ PCA on the pre-selected set of 
candidate cotrending time series and use only those PCA components 
that are `essential', based on their eigenvalues, a product of the 
PCA analysis. From the very first application in the SysRem algorithm 
by \cite{tamuz2005}, this selection method has been widely used, 
often supplemented by other criteria, such as selecting dominant 
components from the PCA set based on the statistical properties of 
the regression coefficients of the cotrending vectors \citep{petigura2012}. 
Or, in another approach, application of maximum likelihood criterion 
by considering the effect of the non-Gaussian nature of the full model 
with systematics 
\citep[i.e., PDC-MAP, the Kepler pipeline by][]{smith2012,stumpe2012}. 

Because the Kepler data are dominated by systematics, rather than 
white noise (at least down to the Neptune/Super Earth regime), here we recognize the importance of avoiding variables 
in the TFA templates set. However, we still do not use any PCA-like 
combination to select the `most dominant' correcting vectors. 
This is because templates, that are `less common' could also be 
useful in the case a some targets. For example, outlier data points 
may lead to the rejection of a particular template in the PCA 
classification, whereas the same template may become rather instrumental 
in filtering out these outlying points if and when they occur in 
a target of interest. This is very important, since the signal of 
interest (the transit) is also some sort of outlier. Preserving 
the transit but treating the outliers `naturally' (i.e., without 
clipping them) is an important ingredient of a successful transit 
(or transient) search. 

\begin{figure}[h]
\centering
\includegraphics[width=0.45\textwidth]{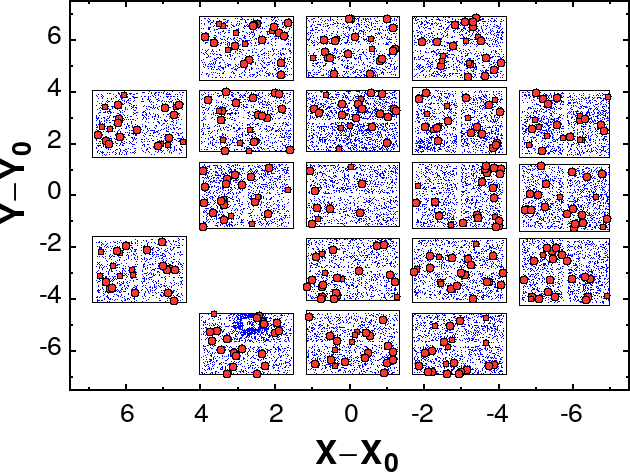}
\caption{Distribution of the TFA template stars (red dots) for 
         $387$ templates. The $X$ and $Y$ coordinates (given 
	 in [deg] relative to the center of the field) result from the 
	 transformation of the equatorial coordinates to remedy strong 
	 distortions of the CCD subfields (see text). The best fitting 
	 rectangles to the hulls of the subfields are shown by black 
	 lines. The small blue dots show the stars with Kepler photometry.}
\label{xy_tfa}
\end{figure}

The main steps in selecting the TFA template set are as follows: 
\begin{itemize}
\item
{\em Spatially uniform sampling:} 
Because systematics are different across the field of view (FOV), 
we aim for spatially homogeneous sampling. From each of the $19$ 
tiles (built up from $2\times2$ closely spaced CCD units) we 
select the same number of templates. Because of the distortion 
introduced by the use of equatorial coordinates $(\alpha,\delta)$, 
we shifted, rotated and stretched/squeezed the FOV so that all tiles 
became approximately of the shape of a rectangle. This step is 
necessary for the simple bookkeeping of the potential templates 
and their association with a given tile. The transformation 
formulae to the new coordinate system $\bf (X,Y)$ are as follows: 
\begin{eqnarray}
\label{ad2xy}
\Delta \alpha & = & \alpha - \alpha_0 \hskip 1mm ,\\
\Delta \delta & = & \delta - \delta_0 \hskip 1mm ,\\
X             & = & (\Delta \alpha \cos \omega + \Delta \delta \sin \omega)\cos \delta\hskip 1mm ,\\
Y             & = & \Delta \delta \cos \omega - \Delta \alpha \sin \omega \hskip 1mm . 
\end{eqnarray}
We determined the transformation parameters with a simple trial and error method. 
For C05 we obtained: $\alpha_0=130.2$, $\delta_0=16.8$, $\omega=-1^{\circ}$.
In each tile the potential templates were chosen from a uniform random 
distribution on the brightness-ordered list of stars. We allowed up to $200$ 
random tries from each tile and selected those stars that satisfied the criteria 
below.  
\item
{\em Focus on low photon noise:} 
Templates are chosen from the bright side of the magnitude distribution 
of the campaign field (in the case of C05, this limit was set at 
$Kp=13.4$, corresponding the brightest $10000$ stars in the field). 
\item
{\em Avoid near singular cases:}
The distance between two template stars should exceed some value, {\em dmin}, 
limited by the PSF of the Kepler telescope. 
\item
{\em Small overall scatter:}
The RMS of the residuals around the fitted low-order polynomial $\{p\}$ 
to the SAP time series $\{x\}$ cannot be greater than $\sigma_t$, 
i.e., $\sigma(x/p)<\sigma_t$. 
\item
{\em Small Fourier power:}
The Fourier fit $\{f\}$ to the residuals of the polynomial fit 
above should not improve the goodness of fit by more than a 
factor of $r_{\sigma}$, i.e., \hfill\break 
$\sigma(x/p)/\sigma(x/p-f)<r_{\sigma}$.    
\end{itemize}  
We used $2^{nd}$ order polynomials and $50^{th}$ order Fourier series in 
all template sets. These parameters were chosen after various tests, 
including visual inspection of the candidate templates. The other 
parameters resulted from the same procedure, and set at the following 
values: $\sigma_t=0.03$ (in relative flux units, normalized to $1.0$ 
in respect of the average flux of the given star), $r_{\sigma}=1.2$. 
The final set of templates is shown in Figure~\ref{xy_tfa}. 
It is intriguing that the tile in the center contains significantly less 
number of templates than all the other tiles. A brief visual examination 
of the light curves (LCs) for this tile revealed an overall lower level 
of satellite rotation adjustment. As the referee suggested, the different 
behavior of the objects on the central tile could be related to the 
closeness of this tile to the rotation axis of the satellite. However, 
the general drift and the ramping at the start of the campaign remained, 
leading to a poor polynomial fit at low order. Consequently, the Fourier 
fit on the residuals became more significant and that led to the failure 
in satisfying the "small Fourier power" criterion by most of the LCs 
associated with this tile. Although the template number distribution 
among the tiles can be made uniform easily by using a higher order 
polynomial fit, this does not change the detection efficacy in any 
major way (i.e., all detections discussed in this paper remain 
reliable/strong).     
  
%
%
%

%
%
\subsection{Optimizing the Fourier order}
\label{sect:four_opt}
Determination of the Fourier order $N_{\rm FOUR}$ is important, 
because in critical cases (e.g., in the presence of high-frequency 
or high-amplitude stellar flux variations), the shallow transit 
signal can be easily washed out by a too high-order fit or by leaving 
in the high-frequency components because of a too parsimony choice 
of the Fourier order. Due to the complexity of the time series, and 
the specific representation of it via a Fourier series with frequencies 
independent of the actual contributing components, standard methods 
(e.g., Fisher test) for optimized parameter number are not 
straightforward to employ. The method presented here is far from 
being exact. It is largely based on the general pattern of the RMS 
of the Fourier fit as a function of the number of the Fourier 
components (in brief the Fourier order). The scheme and the parameters 
used are derived form the large number of tests performed on the 
Campaign 5 data and serve one goal only: separate the smooth Fourier 
component from the discrete transit events. 

It is also important to investigate if testing the SAP LCs is 
suitable for a reliable estimation of the best Fourier order. 
We found that systematics may lead to serious errors in the estimated 
order. Therefore, we decided to perform a TFA filtering first, and 
then employ the method described below on the filtered data to estimate 
the Fourier order. 

The main goals of the optimization are as follows:  
i) find a simple and easy to use criterion that handles most cases 
of stellar variabilities encountered in the Kepler survey; 
ii) minimize mixing instrumental systematics and stellar variability; 
iii) perform the determination of the optimum order quickly 
(before the complex filtering of the input time series, involving 
systematics and stellar variability);  
iv) aim for low Fourier order (CPU demand increases sharply by 
higher Fourier order).  
   
All these point to seeking low-order fits, and we restrict the search 
to $N_{\rm FOUR} < 150$. Concerning iii), we linearly interpolate the 
TFA-filtered LC of the target to an equidistant basis and perform the 
test by using simple Fourier transforms (equivalent to a least 
squares fit on an equidistant timebase, but much faster). 

As expected, $\sigma(N_{\rm FOUR})$ -- the RMS of the residuals of the fit -- has a very steep decreasing 
part for low $N_{\rm FOUR}$ and then, a long, nearly linearly decreasing 
tail toward higher $N_{\rm FOUR}$ values. If there is a high-frequency 
stellar variation, then we have other steep drops in RMS at large-enough 
$N_{\rm FOUR}$ values. The nearly flat parts in the $\sigma(N_{\rm FOUR})$ 
function make the simple Fisher test difficult to use for parameter 
optimization. Therefore, the method described below considers both 
the statistical properties of the RMS of the Fourier fit and the 
possible late (i.e., high $N_{\rm FOUR}$) convergence of the fit.   

\begin{itemize}
\item
Fit the SAP LC with the TFA template set to be used by the routine 
performing the full analysis for the target. Use the residuals of this 
fit in the subsequent steps. 
\item
Compute $\sigma(N_{\rm FOUR})$ for $N_{\rm FOUR}=1-150$ by omitting $5$\% of the data points at 
both edges (to avoid extra increase in RMS due to the Gibbs phenomenon). 
\item
Perform robust linear fit to $\sigma(N_{\rm FOUR})$ between 
$N_{\rm FOUR}=100$ and $150$. 
\item
Compute $S1$, the RMS of the above linear fit.  
\item
Extrapolate this line all the way to the lowest Fourier order, 
and compute the difference $\Delta L$ between $\sigma(N_{\rm FOUR})$ 
and the extrapolated line $L(N_{\rm FOUR})$.  
\item
The first type of optimum Fourier order $M1$ is defined as the 
largest order at which $\sigma-L$ is greater than $S1$. 
\item 
Compute the ratio $R2(N_{\rm FOUR})=\sigma^2(N_{\rm FOUR})/\sigma_{150}$, 
where $\sigma_{150}$ is the 
standard deviation of $\sigma^2(N_{\rm FOUR}$) at the  
highest Fourier order tested, i.e., 
$\sigma_{150}=\sigma^2(150)\sqrt{2/(0.9n)}$, where $n$ is the 
original number of data points with a factor taking into account 
the data cut at the edges (see above)\footnote{The formula cited 
for the standard deviation of the sample variance is valid under 
the assumption that the residuals follow a Gaussian distribution. 
For an easy reference, see: \tiny{
\url{https://stats.stackexchange.com/questions/29905/reference-for-mathrmvars2-sigma4-left-frac2n-1-frac-kappan
}}}. 
\item
The second type of optimum Fourier order $M2$ is defined as the 
largest order at which $R2(N_{\rm FOUR})-R2(150) > 6$. 
\item
Since we found too low order fits often insufficient, the optimum 
Fourier order is defined as follows: $M_{\rm opt}=MAX(20,MIN(M1,M2)+10)$. 
\end{itemize}

%
%
\begin{figure}[h]
\centering
\includegraphics[width=0.40\textwidth]{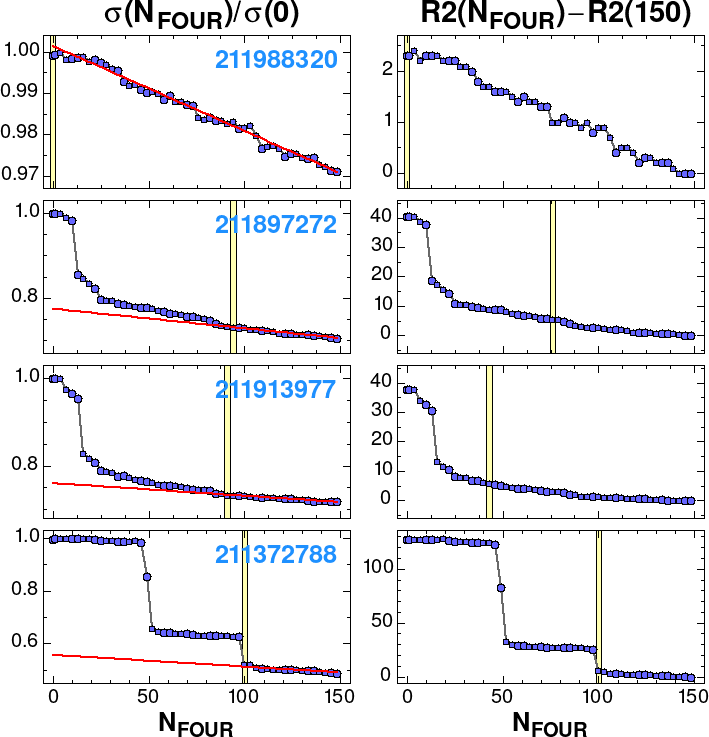}
\caption{Examples on the optimized selection of the Fourier order. 
{\em Left column:} normalized standard deviation of the Fourier 
fit as a function of the Fourier order. Red line shows the robust 
linear regression to the data in $N_{\rm FOUR}=100-150$. 
{\em Right column:} Fit variances over the standard deviation of 
the fit variance at $N_{\rm FOUR}=150$. Vertical yellow stripes show 
the optimum Fourier orders derived from the statistics given in the header. 
See text for further details.}
\label{opt_four}
\end{figure}

Figure~\ref{opt_four} exhibits four typical types of the functions 
described above. Examples shown are from the list of planetary 
candidates of \cite{kruse2019}. The uppermost panel shows the pattern 
for an apparently non-variable star. As expected, the optimum Fourier 
order is low. The $R2$ statistic shows a similar pattern. Since $R2$ 
gives the change in $\sigma^2(N_{\rm FOUR})$ normalized to its theoretical 
error at the highest Fourier order tested, the absolute value of the 
change in $R2$ is important. This is different from the statistic 
derived from the linear fit to the tail of $\sigma(N_{\rm FOUR})$, 
that yields a measure only of the topological property of the RMS 
variation.    

The next panel shows the case when the size of the stellar 
variability is similar to the size of the systematics. The robust 
fit and the interval chosen to perform the linear regression leads 
to a bad approximation of the low-$N_{\rm FOUR}$ regime, reflecting  
the trend expected from the high-$N_{\rm FOUR}$ regime, where the 
harmonic content of the signal is supposed to be exhausted. The 
point where the prediction deviates by more than a given amount 
(see above) yields an estimate of the minimum Fourier order. 
Similarly, the $R2$ statistic measures the relative change, and 
when it exceeds a certain amount, we can consider that value of 
$N_{\rm FOUR}$ as the minimum value needed to represent the Fourier 
content of the signal.   

The other two examples are dominated by stellar variability. The 
target in the bottom panel is a good example for the stepsize 
variation of $\sigma(N_{\rm FOUR})$ and the need to map this 
function to rather high values. From the rule given in last 
step of the optimization process, from top to bottom, the finally 
used Fourier orders are as follows: $20$, $86$, $53$ and $110$. 

%
%
\begin{figure}[h]
\centering
\includegraphics[width=0.45\textwidth]{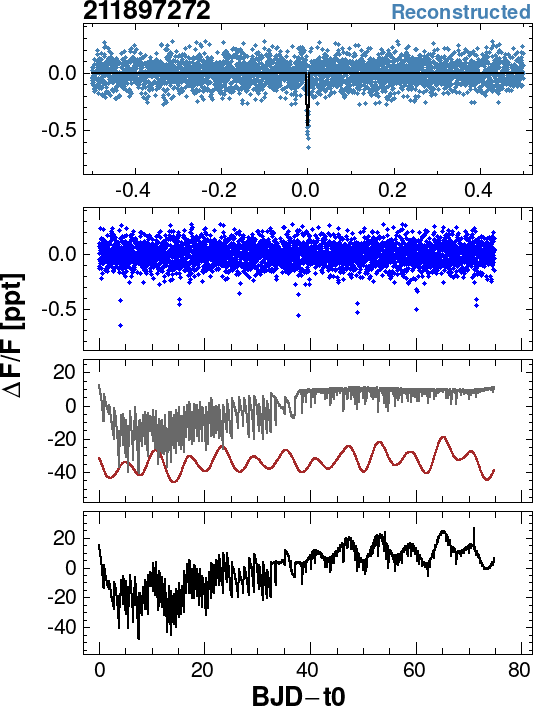}
\caption{Example of the capability of joint robust Fourier and TFA 
fit in disentangling stellar variability and measurement systematics.
The target name and the type of the final data product are indicated 
in the header. The Fourier order is optimized at $N_{\rm FOUR}=83$. 
From the bottom to top: raw (SAP) LC from KEP; 
TFA and FOUR components; raw minus (TFA+FOUR) components -- the periodic 
dips due to the transit signal emerge; folded LC with the best BLS 
period. Continuous black line: best fitting transit signal.}
\label{211897272_diag}
\end{figure}

To illustrate the overall performance of the code and the 
validity of the optimized choice of the Fourier order, 
Fig.~\ref{211897272_diag} shows the case of EPIC 211897272. 
The quasi-periodic stellar variability has been well disentangled 
from the systematics and resulted in the detection of the transit 
signal with a depth of $0.5$~ppt (some 40-times smaller than the 
amplitude of the SAP signal, heavily scrambled by systematics 
in the first half of the campaign). 

In summary, to avoid excessive run times by using flat, overestimated 
Fourier order for all targets (and risk loosing some of the important 
candidates because of overfitting), optimization seems to be an 
important part of the data processing. We found that well over $90$\% 
of the potential candidates can be found in this way and get signatures 
of hidden signals for the remaining few percents. These targets can be 
resolved by performing more detailed test, including detection sensitivity 
against changing the Fourier order.

%
\section{Significance of the signal reconstruction (complete signal modeling)}
\label{sect:rec}
When unknown signal components are searched without the exact 
knowledge of the perturbing components, 
it is unavoidable to introduce some deformation in the signal 
searched for. This is simply because of the (natural) incompleteness 
of the model we use for signal search. This effect is present 
even by using some ``protective shield'' for the unknown component, 
like the self-adjusted weights on outliers as used in this paper. 
There have been attempts to avoid signal degradation due to the 
use of incomplete signal models \citep{foreman2015,angus2016,taaki2020}, 
but subsequent work by \cite{kovacs2016} showed that full modeling 
in signal search is considerably less efficient than thought, 
and, obviously, more CPU demanding.    

Although the overall accuracy is always a focus of attention 
in deriving parameters of interest in any physical system, certain 
parameters may bear higher importance. In the particular case of 
extrasolar planets, the radius (through the scaleheight) is a 
crucial parameter in planetary atmosphere models, and, it is 
obviously important in population synthesis studies. For example, 
the ``radius gap'' problem \citep[the low number density of planets 
at $R_{\rm p}\sim 2R_{\rm Earth}$ -- see, e.g.][]{vaneylen2018, fulton2018} 
is quite sensitive to the radius error, that should be pushed below 
$5$\% to sharpen the feature needed for further theoretical 
considerations \citep{petigura2020}.

Constituting the last step of the signal search, reconstruction 
requires the knowledge of the signal period and the approximate 
parameters of the transit. By adding this approximation to complete 
the signal model, we can iteratively improve the signal parameters 
by solving this complete model \citep[see][for the first application 
of this approach]{kovacs2005}. Here we use Eq.~\ref{eq:F} supplied 
by the successive approximations of the transit signal to minimize 
Eq.~\ref{eq:D} with the adjusted/optimized weights \{$w(i)$\}. 
We note in passing that full modeling is a must in most cases for 
precise transit parameter determination and used commonly in planet 
atmosphere studies based on space observatory data such as HST and 
Spitzer \citep{knutson2008}.   
  
In the following we investigate the signal preserving properties 
of \textsc{tran\_k\oldstylenums{2}} from injected signal tests 
performed on a subset of field C05. Because of the substantial 
size of the parameter space, we simplify the test to all transit 
parameters fixed, except for the transit depth, our main focus of 
interest. In preparing the tests, first we define our detection 
parameters and criteria.    

%
%
\subsection{Signal detection parameters}
\label{sect:det_par}
We consider the signal detected in a simulation if all criteria 
below are satisfied: 
\begin{itemize}
\item[1.]
$|1-P_{obs}/P_{inj}| < 0.002$
\item[2.]
$\delta < 0$
\item[3.]
$SNR_{sp} > 6$ with $SNR_{sp} = (sp(peak)-<sp>) / \sigma(sp)$
\end{itemize}
Here $P_{obs}$ denotes the observed period at the peak power of 
the BLS spectrum of the time series with the injected test signal 
of period $P_{inj}$. The second condition filters out possible 
flares, whereas the third one is our standard condition to 
characterize the signal to noise ratio (SNR) of the BLS spectrum 
($sp(peak)$, $<sp>$ and $\sigma(sp)$, respectively, stand for the 
peak value, the average and for the standard deviation of the 
spectrum). The search was performed in the frequency band 
$[1/T,2.5]$~c/d, where the lower limit comes from the constraint 
of avoiding large gaps in the trial phase-folded LCs at periods 
longer than the total time span $T$. To minimize low-frequency power 
surplus \citep{bakos2004}, the spectrum was robustly fitted by a 
$6^{th}$-order polynomial and the residuals were used to compute 
$SNR_{sp}$. 

Although in the particular case of test signals we did not use 
any other parameters to characterize the significance of the signal, 
in the transit survey performed on the field of C05, we also utilized 
the quantity we call Spectral Peak Density (SPD). This parameter 
is aimed for the quantification of the sparsity of the spectrum, 
yielding small values for those with few large peaks and large values 
for those with many small ones. The former spectra are more likely 
to contain significant signal, whereas the latter ones look more 
closely to what we expect from a noise. For further reference, 
SPD is defined as follows:
\begin{eqnarray}
\label{spd}
SPD & = & {N(sp/sp(peak)> sp_{cut}) \over N(sp)} \hskip 1mm , 
\end{eqnarray}
Where $N(sp/sp(peak)> sp_{cut})$ is the number of spectral values 
(normalized to the highest peak) exceeding a certain cutoff from 
the available $N(sp)$ spectral points. Trained on the C05 data, we 
found that $sp_{cut}=0.3$ yields a quite reliable estimate of 
the ``cleanliness'' of the spectra and the derived $SPD$ values 
are consonant with the visual inspections.  

Although $SNR_{sp}$ and $SPD$ yield useful information on the 
signal content of the time series, the quality of the derived 
folded light curve may not always entirely in agreement with the 
scores received from these parameters. This is because spectral 
parameters reflect the relative significance of the peak component 
to other possible components. For example, in the case of rare or 
single -- otherwise high-SNR -- events, $SNR_{sp}$ and $SPD$ will, in general, indicate 
low significance, whereas the folded LC will obviously suggest 
the presence of a strong signal. Therefore, largely following 
\cite{kovacs2005}, we characterize the quality of the folded LC 
by the Dip Significance Parameter (DSP): 
\begin{eqnarray}
\label{dsp}
DSP & = & {|\delta| \over \sqrt{var_{\delta}+var_{oot}+var_{diff}}} \hskip 1mm , 
\end{eqnarray}
where $|\delta|$ is the absolute value of the transit depth, $var_{\delta}$ is the variance 
of the average (i.e., square of the error of the mean) of the 
intransit data points (i.e., all points from the first contact to the 
last one). In the evaluation of the remaining variances we divide 
the phase-folded LC into bins of the same length as the full 
transit.\footnote{Choosing the bin size the same as that of the 
full transit ensures similar statistical treatment of all parts 
of the LC.} By omitting the bin corresponding to the transit, we 
compute the bin averages for the $N_b$ out of transit (oot) bins 
\{$a_{oot}(j); j=1,2, ..., N_b$\}. The variance of these values 
is $var_{oot}$. This quantity may not decrease DSP to the level 
needed in the case of ragged LCs with large variations between 
the oot bins. Therefore, we added the average of the squared 
differences between the adjacent bin averages, i.e., 
$var_{diff}={1 \over (N_b-1)}\sum_{j=1}^{N_b-1}(a_{oot}(j+1)-a_{oot}(j))^2$. 

%
%
\subsection{Tests: transit depth and transit duration}
\label{sect:tests}
As we have already mentioned earlier in this section, we limit the 
tests on signals that differ only in transit depth, with all other 
parameters fixed during each simulation. Two types of transit signal 
are used:
\begin{itemize}
\item
Signal A:
$P_{\rm tr}=11.111$~d, $T_{14}=0.2$~d, $T_{12}/T_{14}=0.2$, 
$0.1 < |\delta| < 1.9$~ppt, no sinusoidal component
\item
Signal B:
Signal A with a sinusoidal component; 
period: $7.692$~d, amplitude: $20$~ppt.
\end{itemize}
We note that the center of transit and the phase of the sinusoidal 
component are unimportant in the present context. The transit 
depth $\delta$ is chosen from a uniform distribution. The range 
of transit depth was chosen to roughly match most of the 
values of the candidates in the list of \cite{kruse2019}.   

We used our ``400'' TFA template set of $387$ stars with low/small 
stellar variability from the PET database, and injected the signals 
above. These signals were tested with and without employing the 
reconstruction option. We focus on the transit depth and duration 
in the comparison of the injected and derived parameters.

\begin{figure}[h]
\centering
\includegraphics[width=0.35\textwidth]{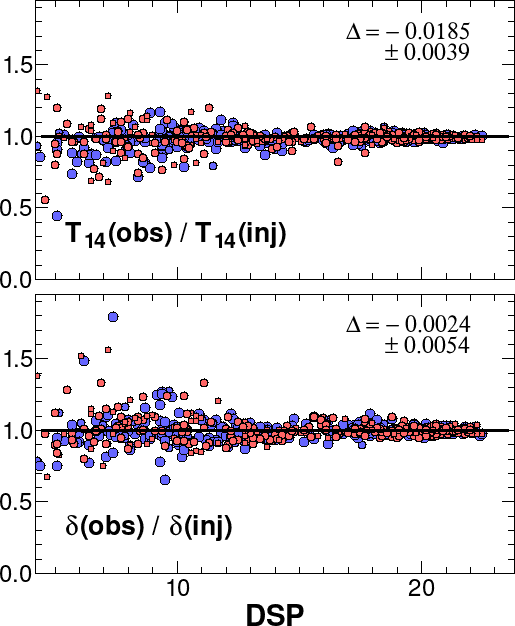}
\caption{Ratio of the observed and injected transit depths $\delta$ 
         and durations $T_{14}$ as functions of the dip 
	 significance parameter DSP. Transit parameters have been 
	 derived by using signal reconstruction. Red and blue points 
	 show the result with (signal B) and without (signal A) 
	 added sinusoidal component to mimic stellar variability. 
	 Inset labels show the mean differences and their errors for signal A. 
	 The estimated transit parameters show no overall bias.}
\label{d_q_rec}
\end{figure}

From the $387$ injected signals of type A, we recovered $297$.  
For type B injections the rate was slightly lower ($294$). 
Figure~\ref{d_q_rec} shows the ratio of the detected and injected 
values for $\delta$ and $T_{14}$. We see that with signal 
reconstruction the derived basic transit parameters are essentially 
unbiased, however, with a scatter increasing toward signals of lower 
significance. Also, except perhaps for the transit duration, there 
is no difference (in statistical sense) between the parameters 
derived from signals with or without a sinusoidal component.     

%
%

\begin{figure}[h]
\centering
\includegraphics[width=0.35\textwidth]{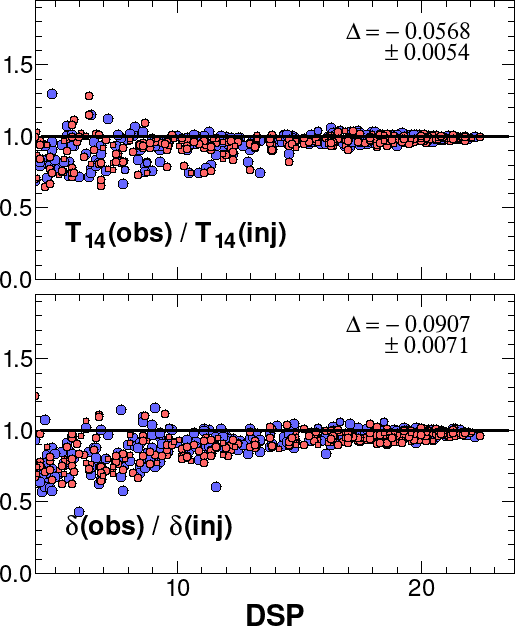}
\caption{As in Fig.~\ref{d_q_rec}, but without employing signal 
         reconstruction. The estimated transit parameters show 
	 significant bias, especially for less significant signals 
	 at lower DSP values.}
\label{d_q_no-rec}
\end{figure}

On the other hand, without employing complete signal model, we 
get a substantial bias (Fig.~\ref{d_q_no-rec}). As expected, 
the observed transit depths are lower than the injected values, 
with a strong increase in this difference toward less significant 
signals. The transit duration follows the same same pattern, albeit 
the bias is somewhat lower.  

%
%

The above tests give supporting evidence that \textsc{tran\_k\oldstylenums{2}} 
is capable of yielding unbiased estimates of the basic transit parameters, 
assuming that a complete signal model is used. We expect that, in 
general, transit depth and duration are estimated better than 
$\sim 10$\% for strong (DSP$>12$) signals. For weaker signals the 
errors increase, but unlikely to go above $\sim 30$\%, while maintaining 
the unbiased nature of the estimates.

%
\section{Comparison with other searches}
\label{sect:comp}
\begin{figure}[h]
\centering
\includegraphics[width=0.40\textwidth]{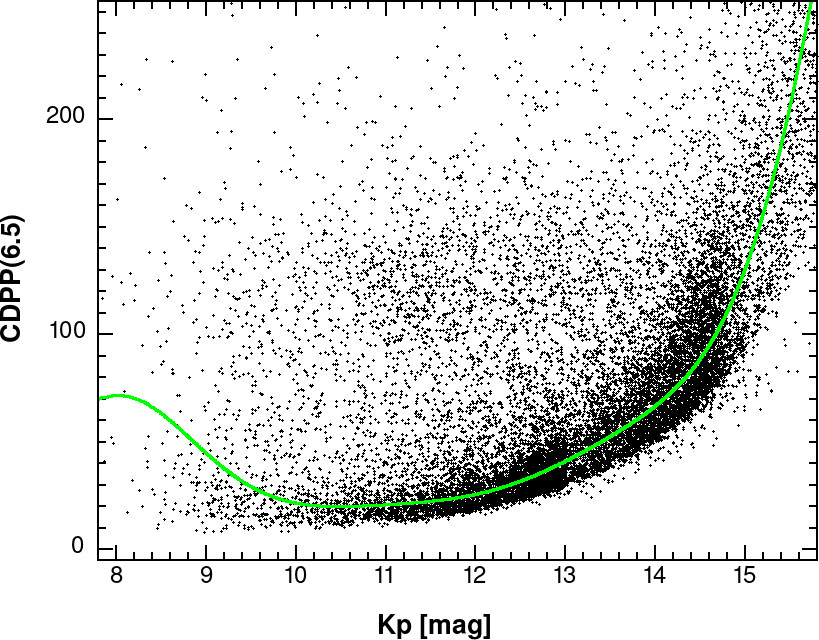}
\caption{Average of the standard deviations of the $6.5$~hour means 
         (in parts per million) vs stellar brightness. The final time 
	 series (i.e., the full, reconstructed models) were used to 
	 calculate the standard deviations from the residuals between 
	 these, and the input (raw) time series. Continuous line shows 
	 the robust fit of a 10-th order polynomial to the data.}	 
\label{cdpp_snr}
\end{figure}

Before we compare the detection rates between our search by 
\textsc{tran\_k\oldstylenums{2}} and other searches, upon the suggestion 
of the referee, we present the averages of the standard deviations of 
the means, taken on a $6.5$~hour timebase of our final data product 
(i.e., including signal reconstruction). This quantity, often called 
as Combined Differential Photometric Precision \citep[CDPP,][]{chris2012}, 
is devoted to sense the potential of transit detection based solely on 
the overall error of the means on a given (transit) time scale. There are 
various approximations to this quantity 
\citep[e.g.,][]{vanderburg2014, aigrain2016}, often including the 
application of a Savitsky-Golay-type (i.e., least squares polynomial) 
filtering of the final data product \citep[e.g.,][]{gilliland2011, luger2018}.      
Here we relied on the full time series model and computed the $6.5$~hour 
CDPP from the unbiased estimate of the standard deviation ($\sigma_{\rm fit}$) 
of the residuals between the model and the input time series. With the 
overall cadence of $0.5$~hours, we get CDPP(6.5)$=\sigma_{\rm fit}/\sqrt{13}$.  
For the $\sim 20000$ stars analyzed, we obtained the result shown in 
Fig.~\ref{cdpp_snr}.  

%
%

\cite{luger2018} reports for the updated EVEREST pipeline basically 
a full agreement for the ridge CDPP values with those of the original 
Kepler mission. \cite{vanderburg2014}, in their Table 1 compare the 
CDPP values of SFF with those of the original Kepler mission. From 
this table we have $18$, $22$, $30$ and $81$~ppm precision at 
$K_p=10.5$, $11.5$, $12.5$ and $14.5$. From the ridge fit to the CDPP 
values in Fig.~\ref{cdpp_snr} we have $20$, $23$, $32$, and $89$~ppm at 
these magnitude values. In spite of the impressive close approximation 
of the precision of the original Kepler mission, we caution that 
low CDPP values tell only that the residuals are small at the particular 
window used, but it does not grant automatically a powerful transit 
detection. This depends on the delicate balance between noise suppression 
and transit signal preservation. 

One of the basic steps in testing the code performance is to verify 
earlier detections and check the consistency of the transit parameters. 
Although this is obviously an important human/machine training ground 
and crucial test in a comparison of the detection efficacy of different 
methods, there are two caveats to keep in mind before making too 
far reaching conclusions from such a test: 
i) the result may depend on the input time series (SAP photometry), 
that are obtained differently by various groups; 
(ii) detection criteria and thoroughness of the analysis might vary 
between the different searches, discarding the trace of transit by 
one search and considering it by other.  

Focusing only on Campaign 5, in the basic verification we relied primarily 
on the $115$ host stars with $138$ planet candidates of \cite{kruse2019}. 
In addition to this, quite recent work, we also tested the $8$ candidates 
in \cite{zink2020}. The basic analysis was performed on the PET database, 
but in unresolved cases we also used the KEP database.    

From the $115$ targets of \cite{kruse2019}, $103$ have been identified 
with high confidence, by using the survey setting of the code (i.e., 
optimized Fourier order with detection criteria listed in 
Sect.~\ref{sect:det_par}). The remaining $12$ targets were examined 
in detail. By using the KEP data, $7$ of these were reclassified 
as ``good/strong'' detections. EPIC 211613886 showed up as a high 
SNR candidate in both datasets, but with half of the period given 
by \cite{kruse2019}. Therefore, we accepted it as a ``detection''. 
EPIC 211988320 shows multiple, apparently non-repetitive transit 
events, yielding high SNR detection, but the favored period is 
different from the one given by \cite{kruse2019}. Because of the 
non-repetitive nature of the events and the significance of events, 
we consider this candidate as ``detected''. The case of EPIC 211939692 
is quite similar, but the Fourier order is overestimated, yielding 
a lower SNR. Detailed examination showed that with lower Fourier 
order this target comes out as strong as EPIC 211988320. Finally, 
we are left with only two candidates (EPIC 211432922 and 211913395) 
that we could not detect, no matter how hard we tried (in agreement 
with the conclusion reached at the various earlier stages while 
developing \textsc{tran\_k\oldstylenums{2}}).  

\begin{figure}[h]
\centering
\includegraphics[width=0.35\textwidth]{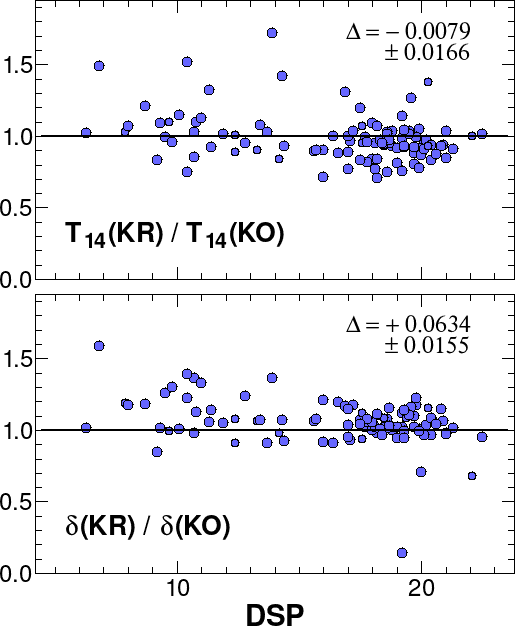}
\caption{Comparison of the transit depth $\delta$ and transit 
         duration $T_{14}$ between this work (KO) and that of 
	 \cite{kruse2019} (KR). The $103$ targets detected from 
	 the analysis of the PET dataset are plotted. The mean 
	 differences and their errors are shown in the upper 
	 right corners.}
\label{d_q_vs_kruse}
\end{figure}

It is an obvious matter of interest to compare the basic transit 
parameters derived in this study and that of \cite{kruse2019}. 
Figure~\ref{d_q_vs_kruse} shows the result of this comparison. 
In the light of the tests presented in Sect.~\ref{sect:tests}, 
it is quite surprising the systematic difference and the large 
scatter we see between the two studies. To add to this discrepancy, 
the injected signal test performed by \cite{zink2020} indicate 
that the EVEREST transit depths are lower by some $2.3$\% as 
they should be. Unfortunately, we have no answer at this moment 
on the source of the apparently significant discrepancy between 
our results and those of \cite{kruse2019}.     

%
%

Interestingly, while testing their planet vetting pipeline on C05, 
\cite{zink2020} found $8$ candidates that do not enter in the list 
of \cite{kruse2019}. At the same time, they could not verify $49$\% 
of the candidates of \cite{kruse2019}. Considering that both studies 
are based on EVEREST data (however, appended with different analysis 
tools and vetting criteria), this is quite intriguing and shows the 
delicacy of planet search in the strongly contaminated environment 
of stellar variability and instrumental systematics. 

In checking the candidates of \cite{zink2020}, we found three targets 
confirmed with high SNR. The $4^{th}$ planet candidate in EPIC 211562654 
is confirmed. For EPIC 211711685 we found a single event in addition 
to the component claimed by \cite{zink2020}. EPIC 212119244 was also 
confirmed, albeit with low SNR at twice of the period given by 
\cite{zink2020}. There remained two candidates (EPIC 211953244 and 
212020330) that we could not confirm, in spite of extensive testing.    

%
\section{Search for new candidates}
\label{sect:new_cand}
Although the field of C05 has been scrutinized by many searches 
\citep{barros2016,pope2016,mayo2018,petigura2018,kruse2019,zink2020}, 
we decided to do the same, based on the high success rate in identifying 
the already known candidates. With the goal of presenting a secure 
list of new candidates for potential followup studies, we surveyed the 
brightest $20000$ stars from the nominally available $\sim 25000$ targets.  
The analysis was run on the KEP and PET datasets separately. 
The signal search was performed in the frequency interval of $[0,3]$~c/d. 
We used the ``400'' template sets for the respective datasets, and the 
optimization method for selecting the Fourier order. 
 
The most viable candidates were selected by using the following 
detection criteria: $SNR_{sp}>7$, $SPD<0.5$, $\delta < 0$, $DSP>5$ , 
N$_{\rm ev}>1$, T$_{12}$/T$_{14}<0.3$ and N$_{\rm ev}$/N$_{\rm int}<0.5$. 
The first two conditions simply require that the BLS spectrum be 
of high-SNR and sparse-enough, as expected from the spectrum of a 
transit signal embedded in moderate to tolerable noise. 
The next two conditions require that the folded signal should 
imply flux decrease, and also be of reasonably significant. The next condition 
makes avoidance of single transits (nevertheless, we found a nice 
one, likely at an earlier stage of search, when such a criterion was 
not employed). To decrease the binary false positive rate, we added 
the condition of minimum steepness of the ingress/egress phase. 
Finally, to avoid false detections due to repeating outliers, we 
required the transit be reasonably well covered by the individual 
events (i.e., the number of transit events should be considerably 
smaller than the number of intransit data points). Clearly, this might 
lead to missing some candidates with transit durations less than an hour, 
or so. 

For the simplest way of linearly decreasing the total running 
time on our multiprocessor computer, the data were analyzed parallel 
in segments containing $5000$ stars in each set. The selection criteria 
above yielded $233$, $127$, $83$ and $39$ items from these sets containing 
objects of decreasing brightness (and, consequently, of higher noise). 
These pre-selected targets were then more deeply inspected and then 
selected as viable candidates if the inspections and tests (e.g., 
stability of the signal against changing the Fourier order, 
sensitivity to the selection of the TFA template set) ended 
up positive. Then, we deselected those candidates that were already 
suggested by the surveys listed above. Thereafter, a brief  
check of the physical plausibility of the detected companion was made 
on the basis of the published stellar parameters from recent 
large-scale studies, aided by the Gaia satellite. The top candidates 
were then further examined for additional transit components. 

We found $15$ candidates\footnote{We could not find traces of these 
candidates in any other earlier publications. However, the single transit 
of EPIC 211503363 is also nicely visible in the processed (PDC, TERRA) 
light curves at the NASA/IPAC exoplanet site.} passing all these steps 
and ending up as potential planetary systems. The derived photometric 
transit parameters for these systems are given in Table~\ref{new_cand}, 
the accompanying diagnostic plots are shown in Appendix A. 

By using the stellar parameters accessible through Gaia DR2 
\citep{gaia2}, the TESS and the K2 stellar catalogs 
\citep{huber2016, stassun2019, hardegree2020}, we checked 
the physical properties of the new candidate systems. As a sanity 
check, assuming central transit and circular orbit, we computed the 
relative transit duration $Q_{\rm tran}=$T$_{14}$/P$_{\rm orb}$ 
through the basic stellar parameters and orbital period: 
\begin{eqnarray}
\label{eq:qtran}
Q_{\rm tran}=0.0756\ R_{\rm s}\ M_{\rm s}^{-{1 \over 3}}\ P_{\rm orb}^{-{2 \over 3}}\hskip 1mm , 
\end{eqnarray}
where the stellar parameters are in solar units, the period is in 
days. Please note that this check does not depend on the blending 
status of the target. 

\begin{figure}[h]
\centering
\includegraphics[width=0.40\textwidth]{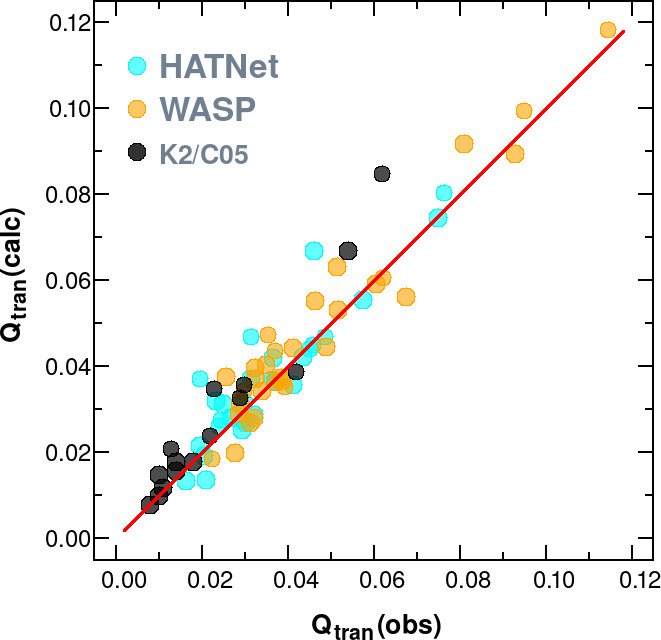}
\caption{Observed vs calculated relative transit durations  
         for the $15$ systems presented in this paper, together with 
	 the $30$ HATNet and $30$ WASP planetary systems detailed in 
	 Appendix~\ref{app_B}. Equation~\ref{eq:qtran} was used to 
	 derive $\rm Q_{\rm tran}(calc)$. The surplus of objects above 
	 the $45^{\circ}$ line indicates the expected effect of off-central 
	 transit. The slight excess of systems below the $45^{\circ}$ line 
	 is due to errors in the stellar masses and radii.}
\label{t14_plot}
\end{figure}

In testing the consistency of the calculated transit durations with the 
observed ones, first we selected systems with well-established planetary status 
(see Appendix~\ref{app_B}, Table~\ref{hj_t14_par}). Then, we plotted the corresponding 
$Q_{\rm tran}$ values together with those of the new candidates. Figure~\ref{t14_plot} 
shows that the new candidates presented in this paper fit well to the overall 
topology of this plot: most of the systems lack exact central transit, therefore, 
the calculated $Q_{\rm tran}$ values are greater than the observed 
values.\footnote{This is also true for the single transiter EPIC 211503363.  
For the minimum orbital period of $\sim 58$~d the calculated transit duration 
is greater than observed ($0.020$ vs $0.013$), allowing longer periods for the 
system.} Several systems have $Q_{\rm tran}(\rm calc) < Q_{\rm tran}(\rm obs)$. 
The most likely cause of this discrepancy is the insufficient accuracy of the 
stellar parameters. Another, albeit less likely/efficient contributing factor   
is orbital eccentricity. The most prominent downward outlier if WASP-4 \citep{wilson2008}. 
All current stellar radius values are in the close neighborhood of $0.90$~$\rm R/R_{\odot}$ 
\citep{bonfanti2020}, making it difficult to increase the calculated $Q_{\rm tran}$ 
value. High resolution imaging \citep{bergfors2013} does not indicate any 
nearby companion that might indirectly affect the transit duration. 

%
%

In the evaluation of the candidates, we estimated the planet radii 
with the assumption that blending was not an issue for any of the targets.  
The stellar and planetary parameters are summarized in Table~\ref{new_can_par} 
of Appendix~\ref{app_B}.  

We found that the majority of the companions have radii between $1$ 
and $2$ Earth radii. We have 5 candidates (EPIC 211537406, 211777794, 
211825799, 211977277 and 211503363) with sub-Jupiter--Neptune radii. 
Except for EPIC 211537406, in this group of more massive planets, 
all companions are around evolved (off of the main sequence) stars. 

The brightest candidate, EPIC 211914045 is special in our sample. 
The star is a K dwarf and the companion has a sub-Earth radius of 
$0.63$~R$_{\rm E}$. The candidate was detected only in the KEP time 
series. This can be the result of the optimized aperture used by the 
KEP pipeline. The PET time series are generated by 
{\sc k\oldstylenums{2}phot} with an aperture size of $3\times3$ pixels. 
The corresponding square covers the target and avoids gathering 
photons from the considerably fainter stars about $10$" to the 
North.\footnote{\url{https://exofop.ipac.caltech.edu/k2/files/211914045/Photometry/\\211914045P-ep20160806.pdf}} 
The exact location and the extent of the optimized aperture of the 
KEP pipeline is not known, but the net flux is higher by some $5$\% 
than the one supplied by {\sc k\oldstylenums{2}phot}. This may indicate 
some level of contamination by the neighbors. Our fully processed LC  
for the KEP data yields RMS$=0.053$~ppt, while for the LC derived 
from the PET data we get $0.061$~ppt. The higher precision is a likely 
contributing factor to the detection in the KEP data. No time series 
are available for the faint visual double to the North, but the nearby, 
similarly bright star EPIC 211913753 to the South, show no signal 
in any of the databases. Because of the lack of full frame images 
with sufficient cadence during the K2 campaigns, we can rely only 
on future dedicated followup works to decide if the faint neighbors 
to the North have any contribution to the signal detected in 
EPIC 211914045.       

%
%
\begin{table*}[h!]
\centering
\begin{minipage}{200mm}
\caption{New planetary candidates from K2 Campaign 5.}
\label{new_cand}
\scalebox{1.00}{
\begin{tabular}{rccrccccccrr}
\hline 
No & EPIC & K$_{\rm p}$ & P$_{\rm orb}\phantom{99}$ & T$_{\rm c}-$T$_{0}$ & $\Delta F/F$ & T$_{14}$/P$_{\rm orb}$ & T$_{12}$/T$_{14}$ & RMS & N$_{\rm dat}$ & N$_{\rm ev}$ & N$_{\rm int}$\\
   &      &    (mag)    &         (day)\phantom{9} &       (BJD)          &   (ppt)      &            -               &            -              &(ppt)&      -        &      -       &  - \\
\hline
 1 & 211914045 & 11.132 &  1.81207 & 40.35521 & $ 0.061\pm 0.006$ & 0.0423 &  0.000 &  0.053 &  3453 & 41 & 147\\
 2 & 211777794 & 12.108 & 19.26493 & 58.84401 & $ 0.239\pm 0.016$ & 0.0183 &  0.113 &  0.091 &  3440 &  4 &  62\\
 3 & 211328600 & 12.568 &  1.04383 & 40.11712 & $ 0.176\pm 0.020$ & 0.0544 &  0.001 &  0.195 &  3445 & 71 & 189\\
 4 & 211528937 & 12.812 &  2.27631 & 40.34633 & $ 0.197\pm 0.016$ & 0.0229 &  0.124 &  0.101 &  3451 & 33 &  78\\
 5 & 211825799 & 12.877 & 33.25434 & 61.12873 & $ 0.563\pm 0.033$ & 0.0139 &  0.006 &  0.148 &  3450 &  2 &  41\\
 6 & 211977277 & 13.037 &  4.80015 & 41.04836 & $ 0.209\pm 0.019$ & 0.0621 &  0.117 &  0.201 &  3620 & 16 & 232\\
 7 & 211503363 & 13.221 & 73.91076 & 97.51803 & $ 4.019\pm 0.047$ & 0.0131 &  0.154 &  0.216 &  3440 &  1 &  43\\
 8 & 212017374 & 14.192 &  9.82573 & 39.91297 & $ 0.332\pm 0.045$ & 0.0142 &  0.003 &  0.231 &  3455 &  8 &  53\\
 9 & 211754117 & 14.199 & 13.79727 & 40.44087 & $ 0.763\pm 0.055$ & 0.0097 &  0.208 &  0.233 &  3445 &  6 &  36\\
10 & 211537406 & 14.565 & 19.43824 & 55.35711 & $ 1.114\pm 0.102$ & 0.0077 &  0.009 &  0.376 &  3429 &  4 &  27\\
11 & 211852237 & 14.594 &  2.68947 & 40.43272 & $ 0.311\pm 0.043$ & 0.0285 &  0.003 &  0.297 &  3445 & 28 &  97\\
12 & 211633000 & 14.601 &  9.23213 & 42.66952 & $ 0.454\pm 0.054$ & 0.0100 &  0.001 &  0.225 &  3449 &  8 &  35\\
13 & 211330455 & 15.148 &  1.62210 & 40.22541 & $ 0.619\pm 0.084$ & 0.0301 &  0.000 &  0.598 &  3427 & 45 & 101\\
14 & 211755530 & 15.279 &  1.74985 & 40.48744 & $ 1.202\pm 0.158$ & 0.0221 &  0.001 &  0.986 &  3455 & 43 &  78\\
15 & 211327678 & 15.540 &  7.71791 & 42.87519 & $ 1.525\pm 0.175$ & 0.0109 &  0.010 &  0.774 &  3445 & 10 &  39\\
\hline
\end{tabular}}
\end{minipage}
\begin{flushleft}
{\bf Notes.} K$_{\rm p}$, P$_{\rm orb}$ and T$_{\rm c}$, respectively, 
stand for the Kepler photometric magnitude, the orbital period and for 
the moment of the center of the transit.  
T$_0=2457100.0$, $\Delta F/F$ denotes the relative flux decrease, 
T$_{14}$ and T$_{12}$ are, respectively, the complete transit and ingress 
durations. RMS stands for the standard deviation of the residuals after 
subtracting the complete time series model, N$_{\rm dat}$, N$_{\rm ev}$ and 
N$_{\rm int}$, respectively, are for the number of data points, number of 
transit events and the number of intransit points. The KEP dataset is used 
for all candidates, except for EPIC 211977277, where we used the PET dataset. 
The period for the single transiter EPIC 211503363 is the formal BLS period, 
close to the total time span of the data. Errors for the transit depths are 
calculated from $e(\Delta F/F)={\rm RMS}\sqrt{2/{\rm N_{\rm int}}}$ 
\citep[see][]{kovacs2019}. 
\end{flushleft}
\end{table*}

We searched for additional planets both among the $115$ systems of 
\cite{kruse2019} and the newly found $15$ systems. As described in 
Sect.~\ref{sect:tran_k2}, we successively subtracted the already found 
transit component, and searched for the next one until the SNR of the 
spectrum reached the noise level. In the final solution all transit 
components were included together with the stellar variation and 
instrumental systematics. The transit depths were fitted simultaneously, 
with the other transit parameters fixed. Naturally, the search 
for additional components is also sensitive to secondary eclipses and 
therefore, it helps in filtering out false positives. We did not find 
any sign for blended binaries among the candidate systems. 

Altogether we found 5 systems that were classified either as single or 
of lower multiplicity. Somewhat more detailed description of the systems 
follows below. The parameters of transit components are given in 
Table~\ref{new_mult}.  

\begin{figure}[h]
\centering
\includegraphics[width=0.40\textwidth]{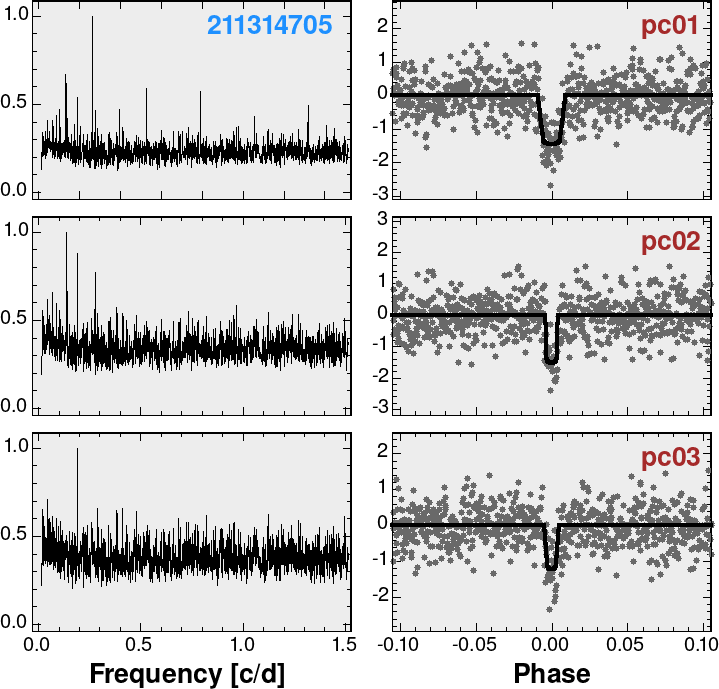}
\caption{The 3-planet candidate EPIC 211314705. {\em Left panels:} 
         successively prewhitened BLS spectra, normalized to unity at 
	 each panel. The spectra are divided into 2000 bins and only 
	 the maxima are shown in each bin. {\em Right panels:} folded 
	 LCs for each component, zoomed on the transit. Continuous 
	 line shows the fit to our transit model. Relative fluxes 
	 are in [ppt].}
\label{mult_211314705}
\end{figure}

{\sl EPIC 211314705}~$-$ This is a 3-planet candidate. No entry in 
\cite{kruse2019}; single planet candidate in \cite{pope2016}. We found 
two additional signals with transit depths $\sim 1.5$ and $1.2$~ppt. 
Somewhat intriguingly, the orbital periods have a constant ratio of 
$1.38$. Figure~\ref{mult_211314705} shows the BLS spectra and the folded 
LCs for the three components.  

\begin{figure}[h]
\centering
\includegraphics[width=0.40\textwidth]{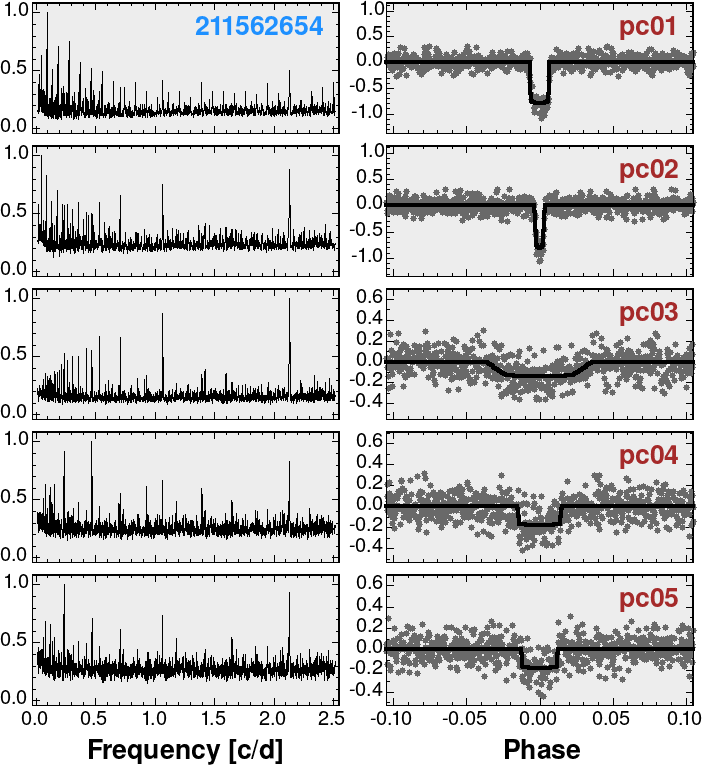}
\caption{The 5-planet candidate EPIC 211562654. Notation is the same 
         as in Fig.~\ref{mult_211314705}. The short period component 
	 pc03 may exhibit TTV, as indicated by the high power remained 
	 at the same frequency in the BLS spectrum of panel pc05 and 
	 the proximity of the associated epoch to that of pc03.}
\label{mult_211562654}
\end{figure}

{\sl EPIC 211562654}~$-$ This is K2-183, a 5-planet candidate. Enters as 
a 3-planet candidate in \cite{kruse2019}. The $4^{\rm th}$ component was 
discovered by \cite{zink2020}. Here we present the $5^{\rm th}$ component, 
in near 2:1 resonance with the $4^{\rm th}$ component. The transit center 
of the $5^{\rm th}$ candidate is located at $T_{\rm c}(4)+0.375P_4$, 
indicating that it is not some sort of artifact of the method used during 
the prewhitening of pc04. The ratio of the relative transit durations is 
not entirely consonant with the expected value from Eq.~\ref{eq:qtran}, 
but is still within the error range indicated by our tests for a signal 
of DSP$\sim 9$ (see Fig.~\ref{d_q_rec}). The $5$ components are displayed 
in Fig.~\ref{mult_211562654}. 

\begin{figure}[h]
\centering
\includegraphics[width=0.40\textwidth]{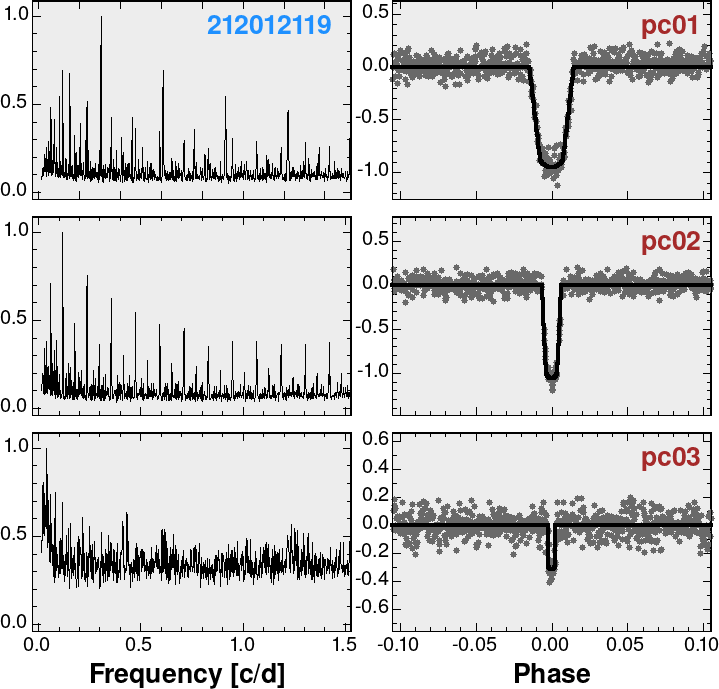}
\caption{The 3-planet candidate EPIC 212012119. Notation is the same 
         as in Fig.~\ref{mult_211314705}. The $3^{\rm rd}$ candidate 
	 is in a close 3:1 resonance with the $2^{\rm nd}$ candidate.}
\label{mult_212012119}
\end{figure}

{\sl EPIC 212012119}~$-$ This is a 3-planet system, listed by both 
\cite{pope2016} and \cite{kruse2019} as a 2-planet system. The $3^{\rm rd}$ 
component is in a near 3:1 resonance with the second component. 
The transit center of the $3^{\rm rd}$ component is near in the middle 
of the 3-period cycle of the $2^{\rm nd}$ component, i.e., 
$T_{\rm c}(2)+1.482P_2$. The observed decrease in the transit length 
is greater than expected from Eq.~\ref{eq:qtran}, but again, this is 
within the error limits. The three components are displayed 
in Fig.~\ref{mult_212012119}.

\begin{figure}[h]
\centering
\includegraphics[width=0.40\textwidth]{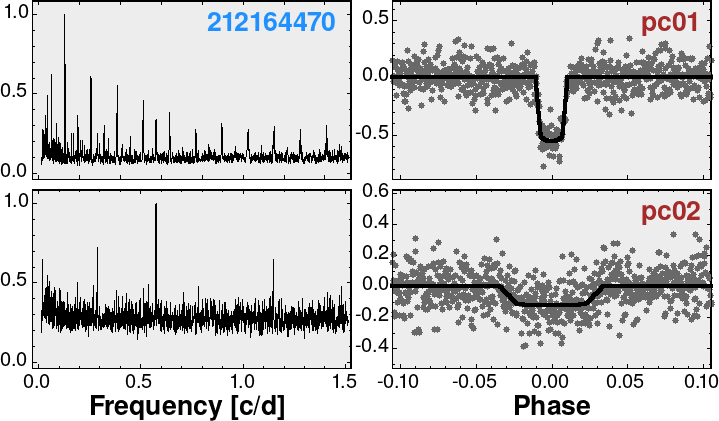}
\caption{The 2-planet candidate EPIC 212164470. Notation is the same 
         as in Fig.~\ref{mult_211314705}.}
\label{mult_212164470}
\end{figure}

{\sl EPIC 212164470}~$-$ This is a 2-planet system, listed earlier 
as a single component system by \cite{pope2016} and \cite{kruse2019}.   
The $2$ components are displayed in Fig.~\ref{mult_212164470}.

\begin{figure}[h]
\centering
\includegraphics[width=0.40\textwidth]{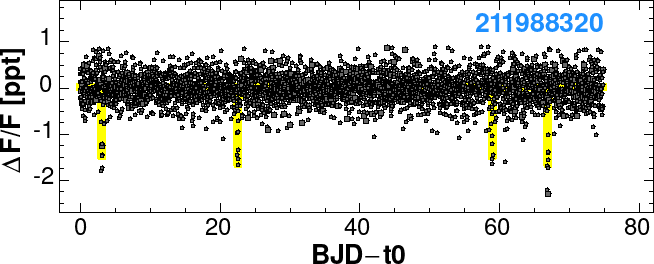}
\caption{Time series of the multiplanet candidate EPIC 211988320. 
         The orbital periods are not well constrained, because of the 
	 rareness of the transit events (shown with yellow background 
	 shading for better visibility. The time series starts at 
	 ${\rm t0}=2457139.6111$.}
\label{mult_ts_211988320}
\end{figure}

{\sl EPIC 211988320}~$-$ This is a likely multiplanetary system, but 
the orbital periods are not well constrained, because of the rareness 
of the transit events. \cite{kruse2019} list this target as a single 
planet system, but as is shown in Fig.~\ref{mult_ts_211988320}, the 
system must host more than one planet. 
%

%
%
%

%
%
%

%
%
%

%
%
%

%
%
%

%
%
\begin{table}[h!]
\centering
\begin{minipage}{100mm}
\caption{New planetary candidates in multiple systems from K2 C05.}
\label{new_mult}
\scalebox{0.95}{
\begin{tabular}{cccccc}
\hline
\multicolumn{6}{c}{EPIC 211314705} \\
\hline
No. &  $f_{\rm orb}$  & $\Delta F/F$  &  T$_{14}$/P$_{\rm orb}$ &  T$_{12}$/T$_{14}$  & T$_{\rm c}-$T$_{0}$ \\
\hline
1 & 0.263623 & 1.464 & 0.0182 & 0.192 & 40.22765 \\
\rowcolor{lightcyan} 2 & 0.138210 & 1.524 & 0.0081 & 0.083 & 42.40394 \\
\rowcolor{lightcyan} 3 & 0.191368 & 1.231 & 0.0101 & 0.231 & 42.43310 \\
\hline
\multicolumn{6}{c}{EPIC 211562654} \\
\hline
1 & 0.092649 & 0.800 & 0.0135 & 0.056 & 47.77446 \\
2 & 0.044202 & 0.804 & 0.0075 & 0.206 & 44.16401 \\
3 & 2.130636 & 0.139 & 0.0735 & 0.185 & 39.64525 \\
4 & 0.464426 & 0.178 & 0.0298 & 0.029 & 40.46272 \\
\rowcolor{lightcyan} 5 & 0.235254 & 0.176 & 0.0248 & 0.019 & 41.26941 \\
\hline
\multicolumn{6}{c}{EPIC 212012119} \\
\hline
1 & 0.304792 & 0.942 & 0.0291 & 0.245 & 42.13420 \\
2 & 0.118495 & 1.054 & 0.0135 & 0.226 & 42.48517 \\
\rowcolor{lightcyan} 3 & 0.039278 & 0.355 & 0.0049 & 0.010 & 54.99590 \\
\hline
\multicolumn{6}{c}{EPIC 212164470} \\
\hline
1 & 0.128066 & 0.556 & 0.0209 & 0.152 & 44.86411 \\
\rowcolor{lightcyan} 2 & 0.573654 & 0.123 & 0.0681 & 0.160 & 40.46965 \\
\hline
\end{tabular}}
\end{minipage}
\begin{flushleft}
{\bf Notes.} 
$f_{\rm orb}$ is the orbital frequency, $\Delta F/F$ denotes the relative 
flux decrease -- in [ppt], T$_{14}$ and T$_{12}$ are, respectively, the 
complete transit and ingress durations, T$_0=2457100$. New planetary 
candidates found in this work are displayed in the colored rows. See text 
for additional details on these systems.   
\end{flushleft}
\end{table}
%

%
%
\section{Conclusions}
In an effort to test some simple ideas in further improving our capability 
to detect shallow transits in the environment of dominating colored noise, 
we developed \textsc{tran\_k\oldstylenums{2}}, a stand-alone Fortran code. 
Development and debugging were performed on the Campaign 5 field data of 
the K2 mission, including the direct use of these data together with various 
test signals injected into the original data. The input data were the raw 
(simple aperture photometry) fluxes available in public archives. 

The input signal is assumed to be built up from four components:  
(i) instrumental systematics, 
(ii) stellar variability,
(iii) white noise  
(iv) transit signal. 
In filtering out (i) we used cotrending (TFA), based on low-variability 
stars in the field and some image property parameters. To eliminate the 
effect of (ii), Fourier modeling was used.

The prime goal was to protect the underlying transit signal from being 
crushed during the pre-BLS phase, and thereby to ensure maximum SNR of the 
resulting frequency spectrum. To reach this goal, we employed the following 
ingredients in the pre-BLS data preparation. 
\begin{itemize} 
\item 
We used general Fourier series representation of the stellar variability 
(i.e., without knowing the particular signal frequencies entering in the 
few-parameter model of the signal). 
\item
To minimize the overshoots due to the Gibbs phenomenon at the edges of the 
dataset, we employed detuned fundamental frequency in the Fourier series 
and autoregressive modeling near the edges. 
\item
To protect the sharp transit features, data adapted weighting was used in 
the robust least squares fit of the combination of the Fourier and TFA parts  
of the model. No outliers were selected at this stage. 
\item
After the above TFA+Fourier filtering, the data were corrected for single 
outliers and flares before passing them to the BLS search. 
\end{itemize} 
Except for single events, the BLS spectrum represents the basic 
statistic that determine if the signal found is interesting or not. 
Although the SNR of the peak remains the prime indicator of the signal 
content, we found useful to introduce the spectral peak density (SPD), 
that characterizes the sparseness of the spectrum.  

A final step in the signal analysis is to consider all signal constituents, 
including the transit components found in the BLS search. In this grand 
robust fit we get a considerable improvement in the precision of the transit 
depths, a vital parameter in planet characterization and prone to underestimation 
when the modeling is incomplete. 

We found that \textsc{tran\_k\oldstylenums{2}} is capable to detect nearly 
all previously claimed systems and planet components, with a missing rate 
of $1-2$\%. We found that the strength of the detections depends at a 
non-negligible level on the input data used. In particular, the database 
from the standard Kepler pipeline and the one produced by {\sc k\oldstylenums{2}phot} often  
result in spectra of different quality, and 
therefore, occasionally, missing candidates in one of these datasets. 

In spite of the several earlier visits of Campaign 5 by various groups, 
we found $15$ new candidate systems and $5$ additional planets in already 
known planetary systems. A brief check made on the physical size of the 
new planet candidates indicate that most of them have radii less than 
2~R$_{\rm E}$ and there is one candidate with R$_{\rm p}=0.6$~R$_{\rm E}$.

\begin{acknowledgements}
This project would not be possible without our new powerful servers. 
We are indebted to the IT management of the Observatory (including 
No\'emi Harnos, Mih\'aly V\'aradi and Evelin B\'anyai) for the careful 
installation and the prompt (and positive) response to our requests. 
Waqas Bhatti from the Department of Astrophysical Sciences of the 
Princeton University is gratefully acknowledged for the essential 
pieces of advice in matters concerning these servers.  
This paper includes data collected by the Kepler mission. Funding for 
the Kepler mission is provided by the NASA Science Mission directorate. 
This research has made use of the VizieR catalogue access tool, CDS, 
Strasbourg, France (DOI: 10.26093/cds/vizier). Supports from the National 
Research, Development and Innovation Office (grants K~129249 and NN~129075) 
are acknowledged. 
\end{acknowledgements}

\begin{appendix}
%
%
\section{New candidates}
\label{app_A}
We show the diagnostic plots for the $15$ new planetary candidate systems 
presented in Sect.~\ref{sect:new_cand}. All candidates are in the Campaign 5 
field of the K2 mission. Please consult with Sect.~\ref{sect:new_cand} 
(Table~\ref{new_cand}) and Appendix~\ref{app_B} (Table~\ref{new_can_par}) 
for details of system parameters and physical properties.  

\begin{figure}[h]
\centering
\includegraphics[width=0.45\textwidth]{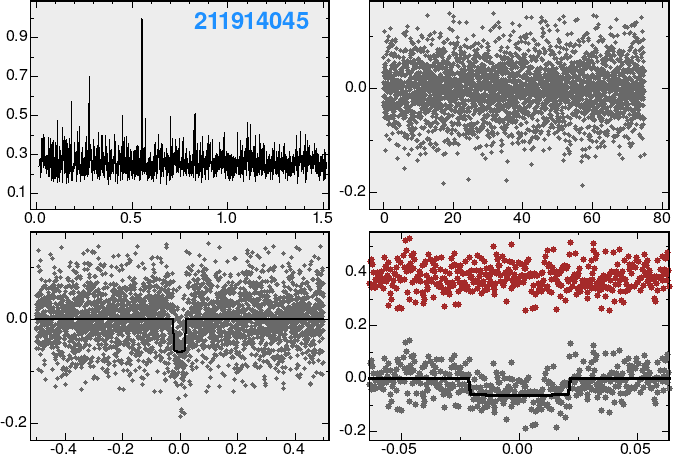}
\vskip 2mm
\includegraphics[width=0.45\textwidth]{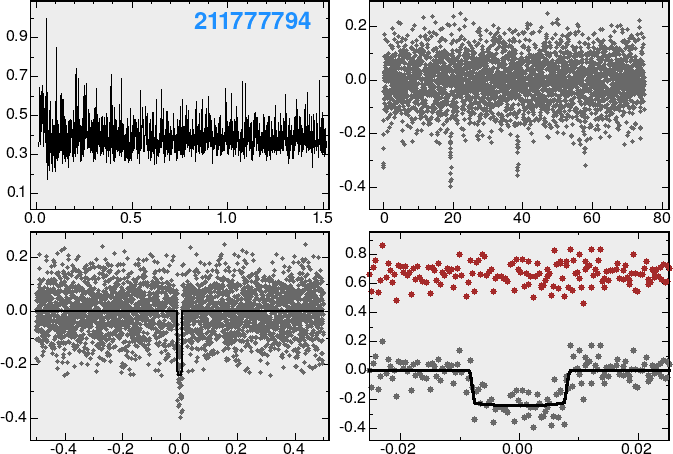}
\vskip 2mm
\includegraphics[width=0.45\textwidth]{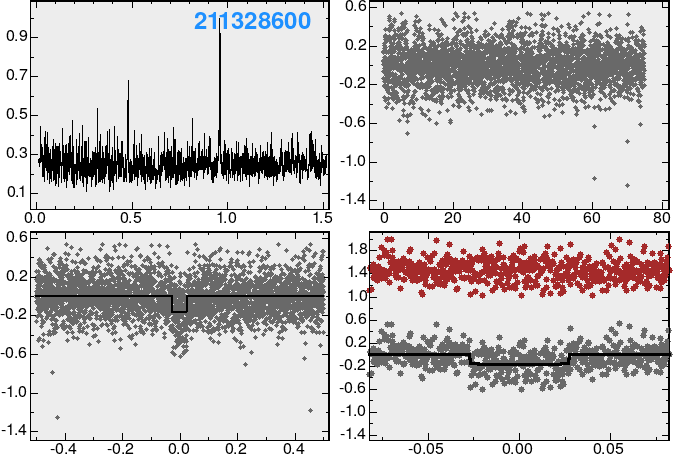}
\caption{Diagnostic plots for the new candidates. Clock-wise 
         for each star: BLS spectrum, time series, zoomed and 
	 full phase-folded light curves. For the zoomed light 
	 curve the residuals (data minus model) are shifted 
	 upward for better visibility. Transit model: black line. 
	 Y axis units: arbitrary (BLS), ppt (others). X axis units: 
	 d$^{-1}$ (BLS), BJD$-2457139.610425$ (time series), phase 
	 (others).}
\label{cand01-03}
\end{figure}
\begin{figure}[t]
\centering
\includegraphics[width=0.45\textwidth]{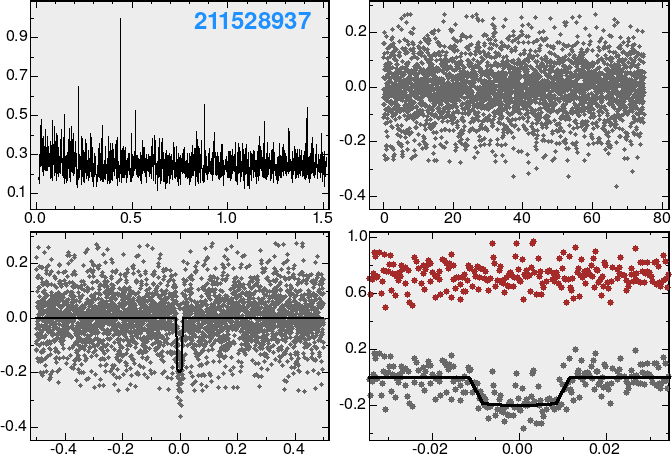}
\vskip 2mm
\includegraphics[width=0.45\textwidth]{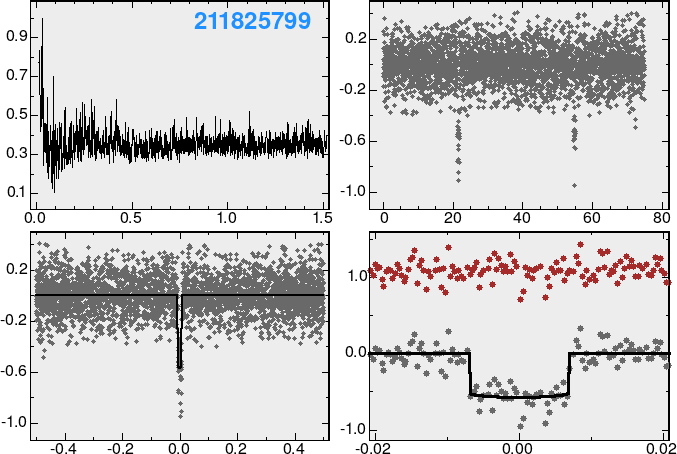}
\vskip 2mm
\includegraphics[width=0.45\textwidth]{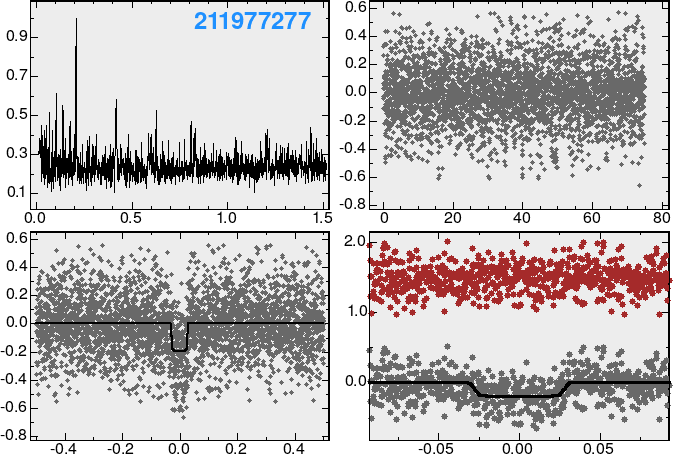}
\caption{Diagnostic plots for the new candidates. See Fig.~\ref{cand01-03} 
         for details. Candidates no. 4 to 6 from Table~\ref{new_cand} 
	 are plotted.}
\label{cand04-06}
\end{figure}
\begin{figure}[t]
\centering
\includegraphics[width=0.45\textwidth]{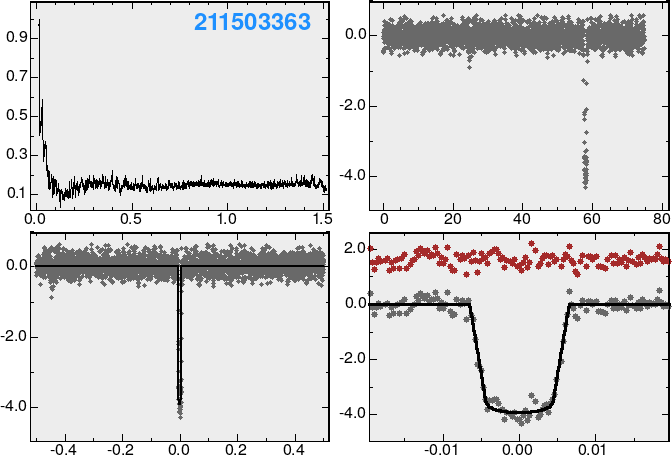}
\vskip 2mm
\includegraphics[width=0.45\textwidth]{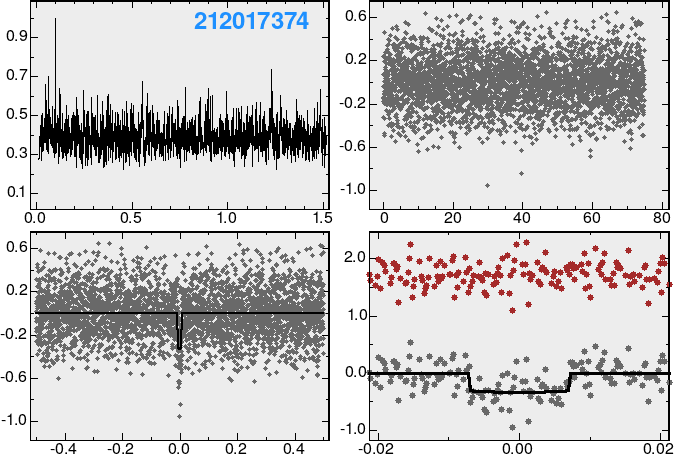}
\vskip 2mm
\includegraphics[width=0.45\textwidth]{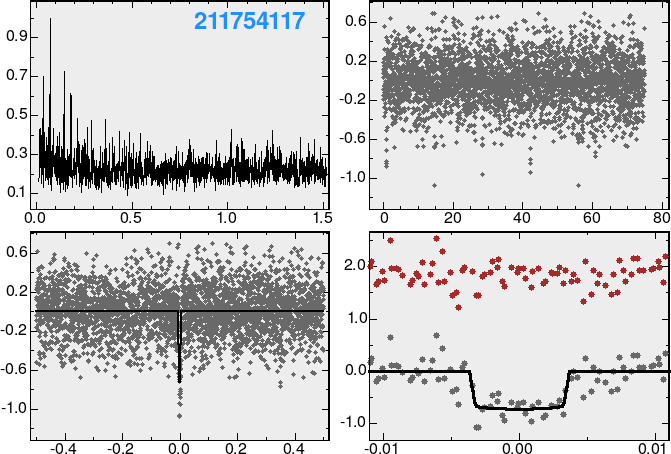}
\caption{Diagnostic plots for the new candidates. See Fig.~\ref{cand01-03} 
         for details.Candidates no. 7 to 9 from Table~\ref{new_cand} 
	 are plotted.}
\label{cand07-09}
\end{figure}
\begin{figure}[t]
\centering
\includegraphics[width=0.45\textwidth]{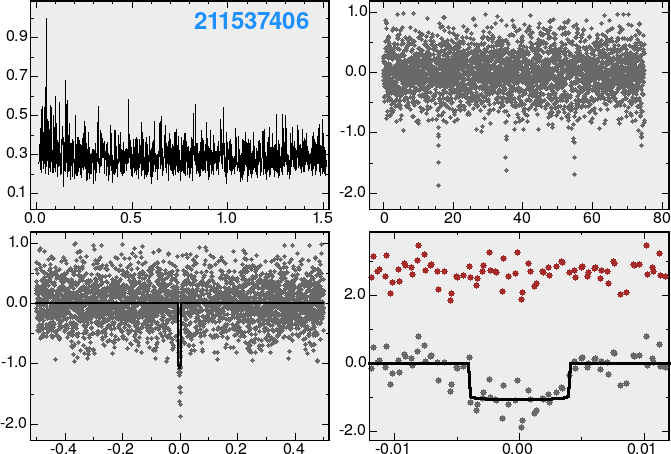}
\vskip 2mm
\includegraphics[width=0.45\textwidth]{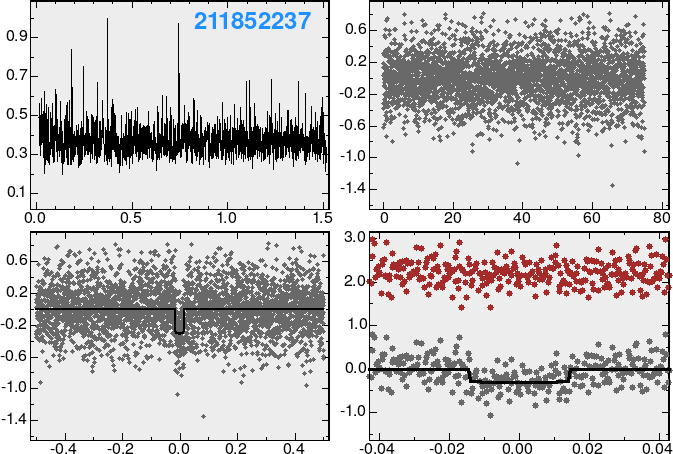}
\vskip 2mm
\includegraphics[width=0.45\textwidth]{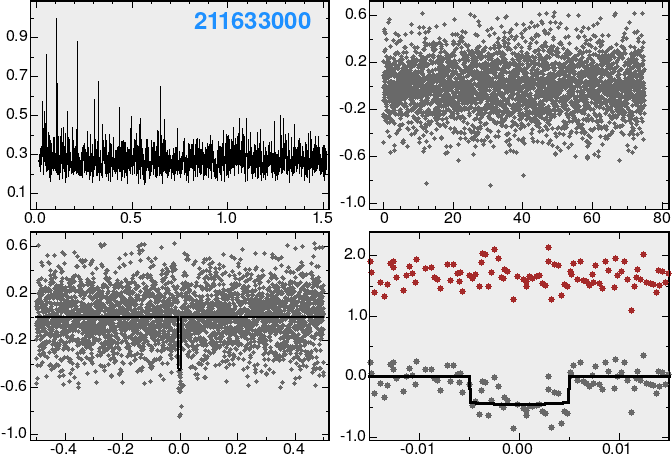}
\caption{Diagnostic plots for the new candidates. See Fig.~\ref{cand01-03} 
         for details.Candidates no. 10 to 12 from Table~\ref{new_cand} 
	 are plotted.}
\label{cand10-12}
\end{figure}
\begin{figure}[t]
\centering
\vskip 2mm
\includegraphics[width=0.45\textwidth]{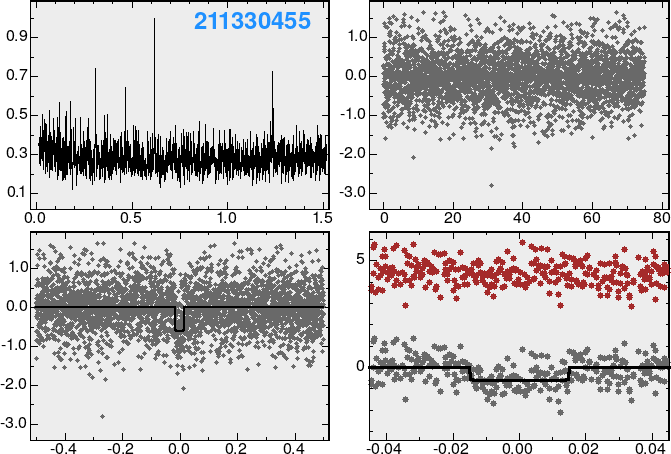}
\vskip 2mm
\includegraphics[width=0.45\textwidth]{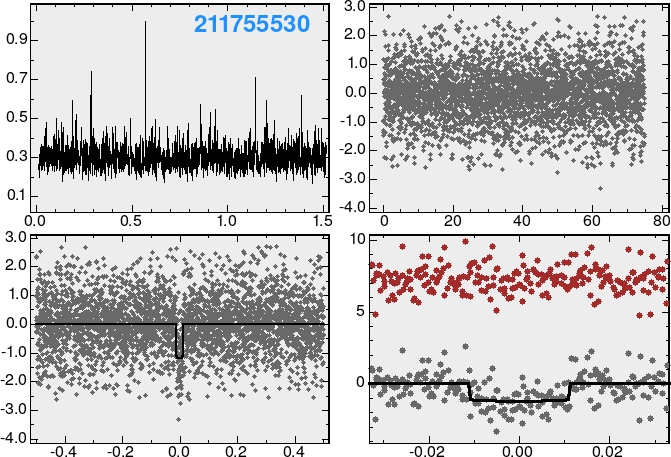}
\includegraphics[width=0.45\textwidth]{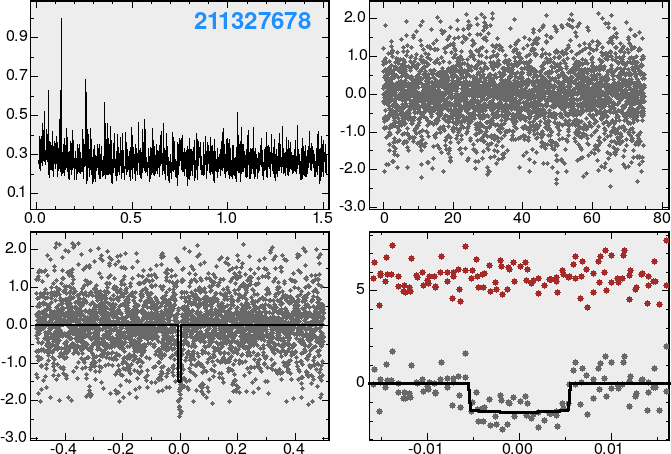}
\caption{Diagnostic plots for the new candidates. See Fig.~\ref{cand01-03} 
         for details.Candidates no. 13 to 15 from Table~\ref{new_cand} 
	 are plotted.}
\label{cand13-15}
\end{figure}
%
%

%
%
\section{Predicted and estimated transit durations}
\label{app_B}
To support Fig.~\ref{t14_plot} here we present the actual values 
used to create the plot. For completeness, we also include the 
coordinates and the planet radii for the new K2/C05 candidates. 

Table~\ref{new_can_par} shows the basic stellar parameters for the 
$15$ new candidates. The stellar parameters are the simple averages 
of the values published by the Gaia Collaboration \citep{gaia2}, 
those given in the K2 and TESS Input Catalogs \citep{huber2016, stassun2019}, 
and the items of a revised K2 catalog by \cite{hardegree2020}. 
Depending on the availability, we could use 2--4 values to calculate 
the averages. For EPIC 211528937 \cite{huber2016} yields a factor of 
two larger radius, therefore, their item is omitted for this 
target. For the single transiter EPIC 211503363 we used P=58~d 
(the minimum period). Please note that this table is aimed primarily 
for the estimation of the transit duration from the stellar parameters 
and orbital period and may not pass the rigor usually followed in the 
analysis of newly discovered individual systems. Nevertheless, the 
consistency of the stellar parameters from various catalogs gives 
enough trust in the estimated transit durations and planetary radii. 

%
%
\begin{table*}[h]
\begin{flushleft}
\begin{minipage}{150mm}
\caption{Some parameters of the new planetary candidates presented in this paper.} 
\label{new_can_par}
\scalebox{1.0}{
\begin{tabular}{cccrccccccc}
\hline
EPIC ID   &      RA    &     DE    & $\rm P_{\rm orb}$ & $\Delta F/F$ & Qobs  & Teff &  <R>  &  <M>  &  $R_p$  &  Qcalc  \\
          & ($^{\circ}$) & ($^{\circ}$) & (d) &  (ppt)  & $-$   &  (K) & (Sun) & (Sun) & (Jup) &  $-$ \\ 
\hline
\hline
211914045 &  129.79767 &  18.93494 &  1.812 & 0.061 & 0.042 & 5154 & 0.734 & 0.883 & 0.057 & 0.039 \\
211777794 &  124.96254 &  17.00097 & 19.265 & 0.239 & 0.018 & 6044 & 1.779 & 1.116 & 0.275 & 0.018 \\
211328600 &  132.13554 &  10.41960 &  1.044 & 0.176 & 0.054 & 5191 & 0.878 & 0.890 & 0.116 & 0.067 \\
211528937 &  127.98521 &  13.54793 &  2.276 & 0.197 & 0.023 & 5016 & 0.769 & 0.898 & 0.109 & 0.035 \\
211825799 &  134.37171 &  17.68293 & 33.254 & 0.563 & 0.014 & 4985 & 2.664 & 1.298 & 0.632 & 0.018 \\
211977277 &  133.91804 &  19.88943 &  4.800 & 0.209 & 0.062 & 4827 & 3.645 & 1.487 & 0.527 & 0.085 \\
211503363 &  127.50679 &  13.18657 & 58.000 & 4.019 & 0.013 & 5121 & 4.098 & 0.996 & 0.260 & 0.021 \\
212017374 &  129.50271 &  20.54328 &  9.826 & 0.332 & 0.014 & 5722 & 0.956 & 0.909 & 0.174 & 0.016 \\
211754117 &  134.14808 &  16.66580 & 13.797 & 0.763 & 0.010 & 5063 & 0.671 & 0.790 & 0.185 & 0.010 \\
211537406 &  135.60571 &  13.66539 & 19.438 & 1.114 & 0.008 & 4380 & 0.657 & 0.694 & 0.219 & 0.008 \\                    
211852237 &  127.06017 &  18.04899 &  2.690 & 0.311 & 0.029 & 4946 & 0.823 & 0.892 & 0.145 & 0.033 \\                
211633000 &  129.29558 &  14.99941 &  9.232 & 0.454 & 0.010 & 5176 & 0.828 & 0.889 & 0.176 & 0.015 \\
211330455 &  130.72883 &  10.45512 &  1.622 & 0.619 & 0.030 & 4031 & 0.547 & 0.569 & 0.136 & 0.036 \\                 
211755530 &  127.25637 &  16.68488 &  1.750 & 1.202 & 0.022 & 3397 & 0.307 & 0.300 & 0.106 & 0.024 \\
211327678 &  130.42308 &  10.40233 &  7.718 & 1.525 & 0.011 & 3818 & 0.460 & 0.455 & 0.180 & 0.012 \\
\hline
\end{tabular}}
\end{minipage}
\begin{flushleft}
{\bf Notes.} 
Equatorial coordinates (RA,DE) refer to J2000. The relative transit duration 
Qcalc was calculated by using Eq.~\ref{eq:qtran}. See text for additional details 
on the data items. 
\end{flushleft}
\end{flushleft}
\end{table*}

Table~\ref{hj_t14_par} contains the main stellar parameters and 
$Q_{\rm tran}$ values for the first $30$ extrasolar planetary systems 
discovered by the two major wide-field ground-based surveys. The stellar 
parameters are from the TESS Input Catalog \citep[TIC, see][]{stassun2019}. 
The transit durations based on TIC have been proven to be more accurate 
than those calculated with the aid of \url{http://exoplanet.eu/} -- 
the calculated $Q_{\rm tran}$ values were less frequently lower than the 
observed values. 

The observed transit durations were gathered from the Exoplanet Transit Database 
\citep[ETD, see][]{poddany2010}. As before, Eq.~\ref{eq:qtran} was used 
to derive Qcalc. 

The sequential numbers of the WASP systems are not continuous, because 
WASP-9 was proven to be a false positive, WASP-11$=$HAT-P-10 and 
WASP-27$=$HAT-P-14. We also note that WASP-30b is not a planetary system, 
but a stellar binary, with a near main sequence F8 primary and a $61$~M$_{\rm J}$ 
secondary, that can be classified as a brown dwarf. Because there is 
no mass entry in TIC for WASP-20, we relied on the analysis of \cite{southworth2020}. 
WASP-20 is a binary, and the stellar parameters correspond to the brighter 
companion.

%
%
\begin{table*}[h]
\begin{flushleft}
\begin{minipage}{150mm}
\caption{Sample of confirmed planets for the transit duration test.}
\label{hj_t14_par}
\scalebox{0.95}{
\begin{tabular}{lrcccccclrccccc}
\hline
     ID   & $\rm P_{\rm orb}$ & Teff &  R    &  M    & Qobs  &  Qcalc &  &  ID   & $\rm P_{\rm orb}$ & Teff &  R    &  M    & Qobs  &  Qcalc \\
          &        (d)        &  (K) & (Sun) & (Sun) &  $-$  &    $-$ &  &       &        (d)        &  (K) & (Sun) & (Sun) &  $-$  &    $-$ \\
\hline
\hline
HAT-P-01 & 4.465 & 5980 & 1.178 & 1.126 & 0.0249 & 0.0316 & |&	WASP-01  & 2.520 & 6200 & 1.564 & 1.153 & 0.0623 & 0.0609 \\
HAT-P-02 & 5.634 & 6414 & 1.704 & 1.290 & 0.0314 & 0.0374 & |&  WASP-02  & 2.152 & 5150 & 0.855 & 0.869 & 0.0349 & 0.0406 \\
HAT-P-03 & 2.900 & 5224 & 0.822 & 0.890 & 0.0298 & 0.0318 & |&	WASP-03  & 1.847 & 6400 & 1.361 & 1.261 & 0.0515 & 0.0633 \\
HAT-P-04 & 3.057 & 5890 & 1.599 & 1.100 & 0.0575 & 0.0556 & |&	WASP-04  & 1.338 & 5500 & 0.895 & 0.970 & 0.0675 & 0.0563 \\
HAT-P-05 & 2.789 & 5960 & 1.130 & 1.051 & 0.0436 & 0.0424 & |&	WASP-05  & 1.628 & 5700 & 1.094 & 1.020 & 0.0606 & 0.0594 \\
HAT-P-06 & 3.853 & 6570 & 1.573 & 1.499 & 0.0366 & 0.0423 & |&	WASP-06  & 3.361 & 5450 & 0.818 & 0.930 & 0.0322 & 0.0282 \\
HAT-P-07 & 2.205 & 6259 & 1.994 & 1.347 & 0.0765 & 0.0806 & |&	WASP-07  & 4.955 & 6400 & 1.469 & 1.360 & 0.0300 & 0.0345 \\
HAT-P-08 & 3.076 & 6200 & 1.431 & 1.285 & 0.0488 & 0.0470 & |&	WASP-08  & 8.159 & 5600 & 0.997 & 0.990 & 0.0225 & 0.0187 \\
HAT-P-09 & 3.923 & 6350 & 1.300 & 1.190 & 0.0365 & 0.0373 & |&	WASP-10  & 3.093 & 4675 & 0.750 & 0.754 & 0.0287 & 0.0293 \\
HAT-P-10 & 3.723 & 4980 & 0.821 & 0.809 & 0.0297 & 0.0277 & |&	WASP-12  & 1.091 & 6360 & 1.749 & 1.170 & 0.1146 & 0.1184 \\
HAT-P-11 & 4.888 & 4780 & 0.760 & 0.770 & 0.0196 & 0.0218 & |&	WASP-13  & 4.353 & 5826 & 1.585 & 1.076 & 0.0372 & 0.0439 \\
HAT-P-12 & 3.213 & 4650 & 0.704 & 0.740 & 0.0303 & 0.0270 & |&	WASP-14  & 2.244 & 6475 & 1.326 & 1.312 & 0.0517 & 0.0534 \\
HAT-P-13 & 2.916 & 5638 & 1.827 & 1.020 & 0.0461 & 0.0672 & |&	WASP-15  & 3.752 & 6300 & 1.503 & 1.176 & 0.0413 & 0.0446 \\
HAT-P-14 & 4.628 & 6600 & 1.545 & 1.433 & 0.0197 & 0.0373 & |&	WASP-16  & 3.119 & 5550 & 1.071 & 1.020 & 0.0256 & 0.0377 \\
HAT-P-15 &10.864 & 5568 & 0.895 & 1.006 & 0.0210 & 0.0138 & |&	WASP-17  & 3.735 & 6650 & 1.573 & 1.354 & 0.0489 & 0.0447 \\
HAT-P-16 & 2.776 & 6158 & 1.242 & 1.164 & 0.0460 & 0.0452 & |&	WASP-18  & 0.942 & 6400 & 1.347 & 1.200 & 0.0951 & 0.0997 \\
HAT-P-17 &10.339 & 5246 & 0.839 & 0.920 & 0.0164 & 0.0137 & |&	WASP-19  & 0.789 & 5500 & 1.028 & 0.970 & 0.0810 & 0.0919 \\
HAT-P-18 & 5.508 & 4870 & 0.740 & 0.773 & 0.0206 & 0.0195 & |&	WASP-20  & 4.900 & 5950 & 1.242 & 1.113 & 0.0289 & 0.0314 \\
HAT-P-19 & 4.009 & 4990 & 0.796 & 0.831 & 0.0294 & 0.0254 & |&	WASP-21  & 4.323 & 5800 & 1.348 & 1.080 & 0.0324 & 0.0374 \\
HAT-P-20 & 2.875 & 4595 & 0.678 & 0.730 & 0.0268 & 0.0282 & |&	WASP-22  & 3.533 & 6000 & 1.191 & 1.170 & 0.0387 & 0.0368 \\
HAT-P-21 & 4.124 & 5588 & 1.256 & 1.000 & 0.0370 & 0.0369 & |&	WASP-23  & 2.944 & 5150 & 0.887 & 0.840 & 0.0340 & 0.0346 \\
HAT-P-22 & 3.212 & 5302 & 1.042 & 0.930 & 0.0372 & 0.0371 & |&	WASP-24  & 2.341 & 6075 & 1.353 & 1.140 & 0.0463 & 0.0555 \\
HAT-P-23 & 1.213 & 5924 & 1.152 & 1.078 & 0.0750 & 0.0747 & |&	WASP-25  & 3.765 & 5750 & 0.877 & 1.025 & 0.0312 & 0.0272 \\
HAT-P-24 & 3.355 & 6373 & 1.419 & 1.250 & 0.0453 & 0.0444 & |&	WASP-26  & 2.757 & 5950 & 1.284 & 1.120 & 0.0355 & 0.0475 \\
HAT-P-25 & 3.653 & 5500 & 0.908 & 0.993 & 0.0321 & 0.0290 & |&	WASP-28  & 3.409 & 6150 & 1.114 & 1.160 & 0.0395 & 0.0354 \\
HAT-P-26 & 4.235 & 5079 & 0.860 & 0.846 & 0.0241 & 0.0263 & |&	WASP-29  & 3.923 & 4800 & 0.636 & 0.880 & 0.0278 & 0.0202 \\
HAT-P-27 & 3.040 & 5300 & 0.865 & 0.916 & 0.0233 & 0.0321 & |&	WASP-30  & 4.157 & 6100 & 1.388 & 1.270 & 0.0384 & 0.0375 \\
HAT-P-28 & 3.257 & 5680 & 1.046 & 1.010 & 0.0412 & 0.0359 & |&	WASP-31  & 3.406 & 6200 & 1.269 & 1.200 & 0.0324 & 0.0399 \\
HAT-P-29 & 5.723 & 6087 & 1.231 & 1.141 & 0.0246 & 0.0278 & |&	WASP-32  & 2.719 & 6100 & 1.028 & 1.300 & 0.0370 & 0.0366 \\
HAT-P-30 & 2.811 & 6250 & 1.340 & 1.254 & 0.0316 & 0.0472 & |&	WASP-33  & 1.220 & 7400 & 1.602 & 1.653 & 0.0928 & 0.0897 \\
\hline
\end{tabular}}
\end{minipage}
\begin{flushleft}
{\bf Notes.} 
Stellar parameter are from TIC, Qobs from ETD, Qcalc from Eq.~\ref{eq:qtran}. 
See text for additional notes. 
\end{flushleft}
\end{flushleft}
\end{table*}

\end{appendix}
\end{document}